\documentclass[fleqn,10pt]{wlscirep}
\usepackage[utf8]{inputenc}
\usepackage[T1]{fontenc}
\usepackage{hyperref}
\usepackage{amsmath}
\usepackage{caption}
\usepackage{epstopdf}
\usepackage{graphicx}
\usepackage[normalem]{ulem}

\newcommand{\bsym}[1]{\boldsymbol{#1}}

\title{Deep learning for diffusion in porous media}

\author[1,*]{Krzysztof M. Graczyk}
\author[1]{Dawid Strzelczyk}
\author[1]{Maciej Matyka}
\affil[1]{Institute of Theoretical Physics,\newline
Faculty of Physics and Astronomy,\newline University of Wroc\l aw, \newline pl. M. Borna 9, 50-204, Wroc\l aw, Poland}

\affil[*]{krzysztof.graczyk@uwr.edu.pl}

\begin{abstract}
We adopt convolutional neural networks (CNN) to predict the basic properties of the porous media. Two different media types are considered: one mimics the sand packings, and the other mimics the systems derived from the extracellular space of biological tissues. 
The Lattice Boltzmann Method is used to obtain the labeled data necessary for performing supervised learning.
We distinguish two tasks. In the first, networks based on the analysis of the system's geometry predict porosity and effective diffusion coefficient. In the second, networks reconstruct the concentration map. In the first task, we propose two types of CNN models: the C-Net and the encoder part of the U-Net. Both networks are modified by adding a self-normalization module [Graczyk \textit{et al.}, Sci Rep 12, 10583 (2022)]. The models predict with reasonable accuracy but only within the data type, they are trained on. For instance, the model trained on sand packings-like samples overshoots or undershoots for biological-like samples.  
In the second task, we propose the usage of the U-Net architecture. It accurately reconstructs the concentration fields. In contrast to the first task, the network trained on one data type works well for the other. For instance, the model trained on sand packings-like samples works perfectly on biological-like samples. Eventually, for both types of the data, we fit exponents in the Archie's law to find tortuosity that is used to describe the dependence of the effective diffusion on porosity.
\end{abstract}
\begin{document}

\maketitle

\section{Introduction}

Diffusion transport in complex porous structures is ubiquitous in nature \cite{Bell61,Shen2007,Kuhn22,Muñoz-Gil2021}. A prominent example is the brain, where diffusion is a dominant process for nutrient and signal transport \cite{Sykova2008,Nicholson2001,Postnikov2022}. The brain's extracellular space, filling the void between neuropil cells, can be treated as a porous medium of complex topology. Effective properties of diffusion in porous media are investigated with various techniques. The application type determines the choice of the method\cite{Tartakovsky2019}. The fields of applications concern diverse domains such as assessment of the tortuosity of extracellular space in brain studies  \cite{Sykova2008, Chen2000}, analysis of the diffusion in Li-ion batteries \cite{Weber2022}, sandy sediments \cite{Shen2007} and studies of hierarchical porous materials \cite{Wernert2022}.

The properties of transport in the void space of porous materials strongly depend on the vastly diverse geometry of its pores. Sandstone rocks, made of grains of the size on the order of tens to hundreds of $\mu \text{m}$ might exhibit pores  multiple times smaller while porosities, defined as a ratio of the pore volume to the overall volume:
\begin{equation}\label{eq:porosity}
    \varphi=\frac{V_\text{pore}}{V},
\end{equation}
can be as low as several percent\cite{Li2016}. On the other hand, the extracellular space, which separates cells in tissues of living organisms, is composed of sheets and tunnels of width on the order of tens of nanometers with numerous dead-ends and cells winding around each other \cite{Kinney2013,Godin2017, Sykova2008}. Therefore, an effort has been made to develop tools to predict diffusive transport properties in porous media of extremely different geometries. In this paper, we adapt convolutional neural networks to study two distinct types of porous media: isotropic granular, corresponding to, e.g., sand packings to rocks, and a system of channels resembling brain extracellular space geometry. 

Diffusive transport is studied on micro and macro scales. In the first, the pore space in which the transport occurs is explicitly modeled. In the second, the effects of the porous medium on the transported molecules are spatially averaged and represented by effective coefficients. 
One distinguishes two general approaches to model diffusion transport. In the first, one traces individual particles (e.g. using random walk). On the other, one considers the concentration field obtained from solving the diffusion equation.
This paper exploits the second approach: we solve the diffusion partial differential equation. We evaluate the concentration field, $c(\vec{r},t)$ ($\vec{r}$ (the position vector, $t$ refers to time), to study the phenomenon. 

The diffusion equation describes spatial and temporal changes in concentration field  of diffusing matter. It can be exploited to model diffusive transport in both scales. At the microscale, for bulk diffusion coefficient $D_0(\vec{r})$ (variable in space), the equation reads:
\begin{equation}
    \frac{\partial c}{\partial t} = \nabla \cdot (D_0(\vec{r})\nabla c).
    \label{eq:diffusion}
\end{equation}
For a constant isotropic diffusivity inside the pore volumes, this simplifies to $\partial c/\partial t = D_0\nabla^2 c$ and the diffusive flux value depends only on the concentration gradient in the direction of the flux.
However, in porous media, the diffusion process is hindered \cite{Tartakovsky08} or accelerated \cite{Kalz22, Alexandre22} due to phenomena like interactions of matter with solid obstructions (steric effects) or binding of species at certain locations in pores. 
In practice we represent the complex porous system by a continuous medium with effective diffusion coefficient $D(t)$, obtained from the Fick's law 
\begin{equation}\label{eq:fick_law}
    \langle D_0 \nabla c(\vec{r},t) \rangle = D(t)\frac{c_\text{out}-c_\text{in}}{L},
\end{equation}
where $\langle \dots \rangle$ denotes averaging over the pore space, $L$ is the distance between the sample inlet and outlet, while $c_\text{in}$ and $c_\text{out}$ are concentrations at the sample inlet and outlet, respectively.
For sufficiently long diffusion times $D(t)$ approaches a steady value $D$, which usually differs from the bulk diffusion coefficient $D_0$. 
The ratio between the bulk diffusion coefficient $D_0$ and the effective diffusion coefficient $D$ is used to define the diffusive tortuosity of a medium:
\begin{equation}
 \lambda=\frac{D_0}{D}.
 \label{eq:effdiffusion}
\end{equation}
The above quantity appears in the scalar transport equation applicable to porous media at the macroscale (see, e.g., Eq. 9 in Sykova and Nicholson\cite{Sykova2008}). The relation between the diffusive tortuosity and porosity of a porous sample is a subject of ongoing debate with a gap to be bridged between experimental and theoretical studies\cite{Shen2007}.

\begin{figure}[t]
\centering
\includegraphics[width=\textwidth]{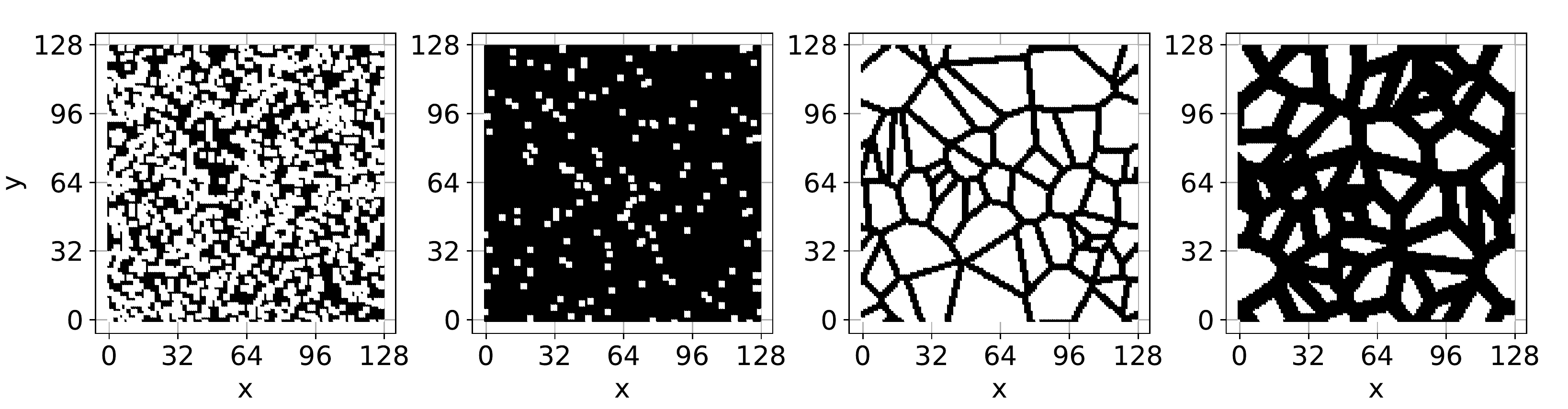}
\caption{Type-A samples with $\varphi=0.42$ and $\varphi=0.92$ (from left: the first and the second, respectively) and type-B samples with $\varphi = 0.32$ and $\varphi=0.66$ (from left: the third and the fourth, respectively). Fluid nodes are denoted by black, while solid nodes are denoted by white. We show samples with low and high porosity values for both data types. \label{Fig:Type_A_and_Type_B_samples}}
\end{figure}

Due to the complexity of the boundary conditions solving analytically Eq.~\eqref{eq:diffusion} is, in most cases, impossible. Numerical solutions have several drawbacks, such as quadratic scaling of timestep or computationally expensive matrix manipulations. 
The above issues become especially prohibitive when one needs to solve the diffusion equation for a large number of data to perform statistical analysis of a given type of porous media, as in Kasthuri \textit{et al.}\cite{Kasthuri2015}.

In this paper, we adopt convolutional neural network (CNN) systems to model the diffusive transport in the porous media. Recently, deep learning-based techniques have been used to compute effective properties of porous media, including permeability, \cite{Kamrava2020, Santos2021}. An example of the usage of CNNs to calculate both permeability and tortuosity is given by Graczyk and Matyka~\cite{Graczyk20}.
In Cawte and Bazylak\cite{Cawte2022}, diffusion and permeability are studied within machine learning tools, including gradient boosting regression, neural network, and support vector regression. Image-based prediction of effective diffusion, $D$, was performed for scanning electron microscope data~\cite{Sethi2022} and random field-based porous media \cite{Roding2022}. Kamrava \textit{et al.}\cite{Karmava2021c} considered the U-Net model, also exploited  in this paper,  to predict the pressure and velocity field's spatial distribution in porous membranes. Permeability of porous samples generated with Voronoi tesselation was predicted within physics informed CNN by Wu et al.\cite{Wu2018}. Kamrava \textit{et al.}\cite{Kamrava2021}  simulated the fluid flow in a porous medium within the deep learning model that incorporated mass conservation and the Navier–Stokes equations in its learning process. Eventually, a comprehensive review of the usage of machine learning methods in studies of  porous media and geosciences is provided by 
Tahmasebi \textit{et al.}\cite{Tahmasebi2020}.


Our goal is to develop a deep learning (DL) system, a deep neural network (DNN), that, based on the system's geometry analysis, predicts either the porosity and effective diffusion coefficient $D$ or concentration field $c(\vec{r})$ in the steady state. We treat both goals separately. In the first case, we consider two types of CNNs architectures. One of them is adapted from our previous work\cite{Graczyk20} on predicting porosity, permeability, and tortuosity in porous media. Additionally, both architectures are modified by adding a self-normalized (SN) module, proposed by Graczyk~\textit{et al.}\cite{Graczyk2022}. The modified architectures work with higher precision than vanilla ones. Moreover, the networks without the SN module tend to overfit the data.
Eventually, we propose to consider the U-Net architecture to predict the concentration distribution \cite{ronneberger2015unet}, and it works with better accuracy and in a broader range than networks that only predict porosity and diffusion coefficient.

The DL systems are obtained from supervised learning. In such an approach, the critical is to work with the data which represent the studied feature comprehensively. We generate two types of samples:  systems of randomly deposited isotropic grains (mimicking the sand packings), and random systems derived from Voronoi tesselation of a plane (mimicking extracellular space of tissues).

In practice, we consider the 2D pictures of the systems with geometry configuration. Then, exploiting the Lattice Boltzmann Method (LBM), we simulate the diffusion process and obtain a steady-state concentration field. From this, using Eq.~\eqref{eq:fick_law}, we calculate the effective diffusion coefficient. The image representing the porous medium's geometry is the input for the network that predicts either the porosity, diffusion coefficient or concentration field. 

The network models are obtained from the statistical analysis of the data, and hence, the networks predict the porosity and diffusion coefficient with some uncertainty. Therefore, we exploit the Monte Carlo dropout technique\cite{gal2015dropout} to estimate how certain in predictions our models are. 

Similar studies were performed by Wu \textit{et al.} \cite{Wu2019}. They adopted AlexNet and ResNet50 CNN models to predict the diffusion coefficient, and they considered one type of data in a broader range of porosity than in our analysis. In contrast, we exploited two other types of porous media. Moreover, our network models are of different architectures. One of the architectures was considered in our previous work\cite{Graczyk20}. The other is the encoder part of the U-Net architecture. All our architectures contain batch normalization layers, which was essential for getting successful fits\cite{Graczyk20}. Eventually, we modify the vanilla models by adding the self-normalization module\cite{Graczyk2022}. Additionally, in this paper, we study the adaptation of the U-Net architecture for reconstructing the concentration field.

The structure of the paper is the following: in Sec.~\ref{Sec:Data}, the method of preparation of the data is shortly presented; in Sec.~\ref{Sec:Sec:DNN:models}, we introduce the network models. The numerical results for both types of the tasks are presented in Sec. \ref{Sec:Sec:Porosity_and_Diffusion}. and  Sec. \ref{Sec:Sec:DNN:Concentration}. We conclude in the  Sec.~\ref{Sec:Summary}. The paper contains Appendix~\ref{App:LBM}, that summarizes  the LBM method.

\section{Diffusion from LBM -- data preparation}

\label{Sec:Data}

\begin{figure}[!ht]
\centering
\includegraphics[width=.8\linewidth]{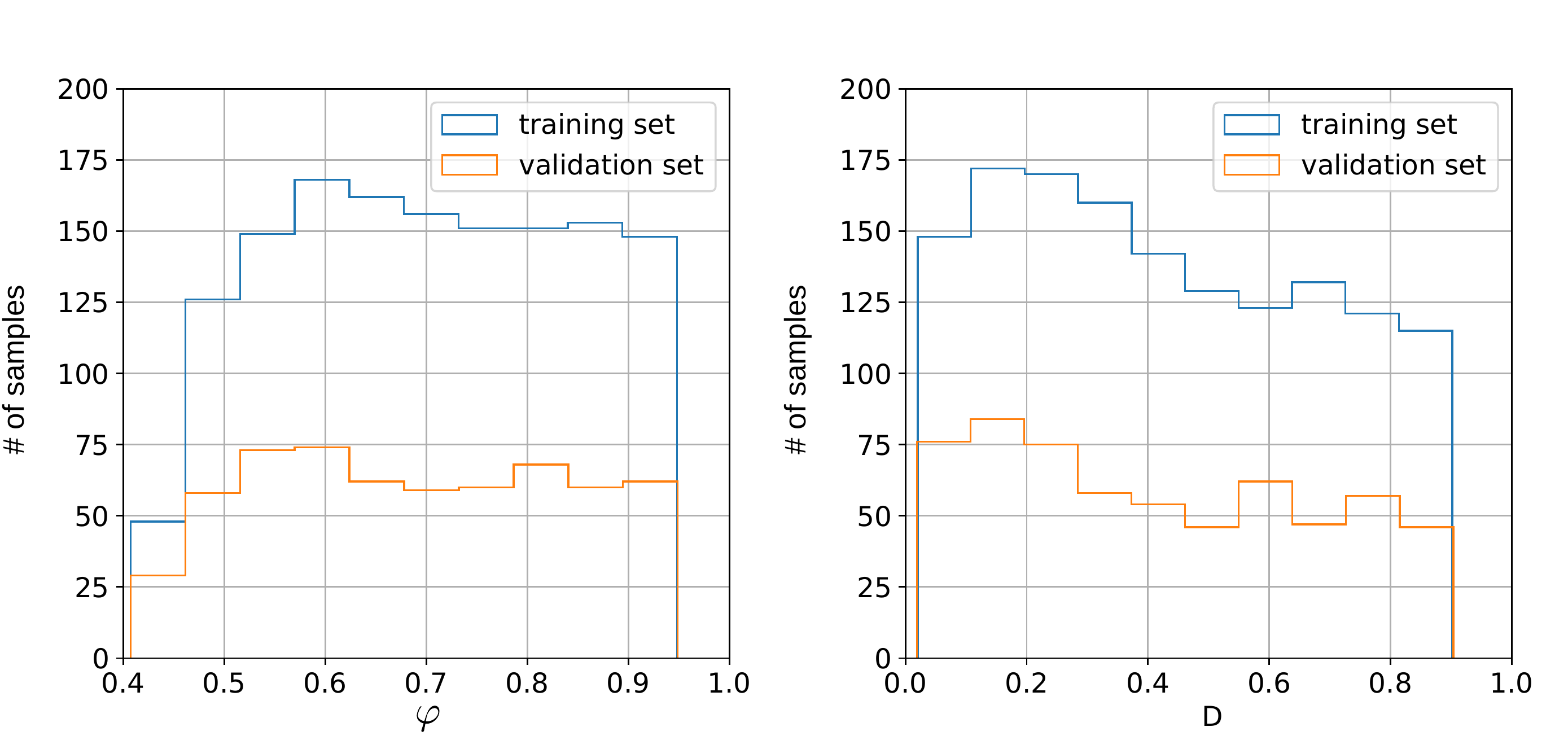}
\caption{Distributions of porosities $\varphi$ and effective diffusion coefficients $D$ obtained with LBM for $2017$ type-A samples ($1412$ and $605$ for training and validation datasets, respectively).}
\label{fig:rect_porous_stat}
\end{figure}
\begin{figure}[!ht]
\centering
\includegraphics[width=.8\linewidth]{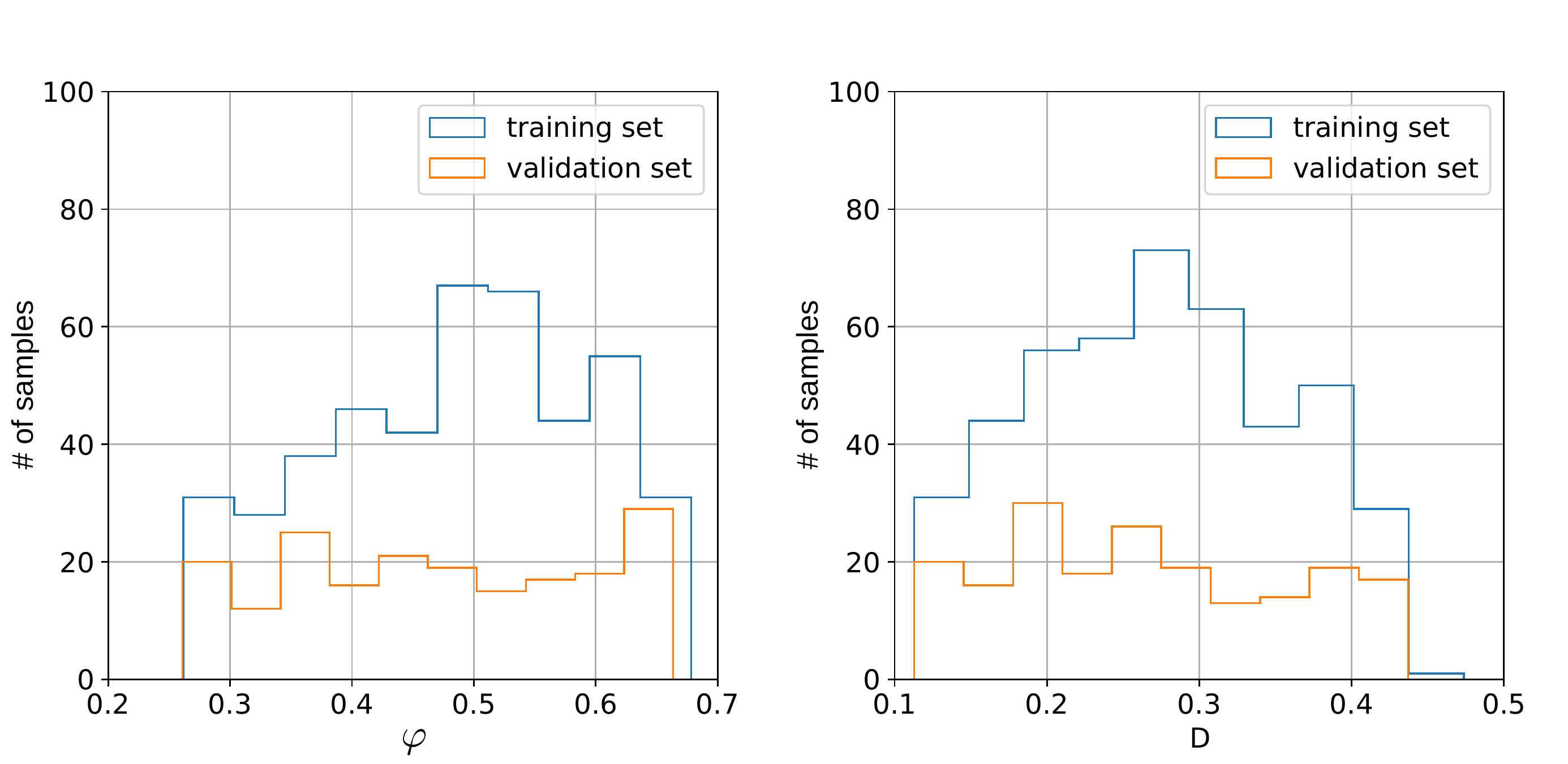}
\caption{Distributions of porosities $\varphi$ and effective diffusion coefficients $D$ obtained with LBM for $640$ type-B samples ($448$ and $192$ for training and validation datasets, respectively).}
\label{fig:voronoi_porous_stat}
\end{figure}

We shall obtain the DL systems that can make predictions about the diffusion phenomenon. To accomplish this task, we need labeled data which will be used to train the DNNs. 
We consider geometries of randomly generated porous domains of two kinds: 
\begin{itemize}
\item type-A: systems of randomly deposited isotropic grains; 
\item type-B: samples based on Voronoi tesselation. 
\end{itemize}
Each sample has a size $L \times L$, $L=128$ nodes. Each node can be either fluid or solid. All fluid nodes belong to the $\mathcal{D}_f$ set, while all solid nodes belong to the $\mathcal{D}_s$ set. The nodes outside the $L \times L$ square belong to the $\mathcal{D}'$ set. 

Both types of systems  have controllable porosity ranging from devoid of obstacles ($\varphi=1$) down to the percolation threshold.
Samples of type-A are idealizations of granular media such as sand packings, clay, or fluidized granular beds. We generate them by placing $3 \times 3$ square obstacles in an initially all-fluid domain; see Fig.~\ref{Fig:Type_A_and_Type_B_samples} (two left panels). The coordinates of obstacle positions are drawn from the uniform distributions in the range $[0,L)$, and obstacles can overlap. Filling the sample configuration with blocks stops when the assumed porosity is achieved. Then the percolation test between the left and right boundary is performed. Samples that do not pass the test are rejected from the dataset.

Type-B domains' geometries are derived from biological systems such as the extracellular space of tissues; see Fig.~\ref{Fig:Type_A_and_Type_B_samples} (two right panels). We generate them by placing $50$ points in $L \times L$ square. The points' position coordinates are drawn from the real uniform distribution on the interval $[0,L)$. Then the set is duplicated to eight neighboring domains to minimize the boundary effects. Then, Voronoi tesselation of $\mathbb{R}^2$ space based on the drawn points is performed to obtain the edges of the polygons, which are the centerlines of the pore channels. We start discretization from all-solid nodes, which corresponded to zero-width channels. Then the width of the channels is iteratively and uniformly increased, and nodes laying inside any channel are marked as fluid, see Fig.~\ref{Fig:Type_A_and_Type_B_samples}. The iteration procedure stops when the assumed porosity is achieved. Due to the nature of the edges network, the samples are guaranteed to percolate from left to right, and no additional tests are performed.

We assume reflecting (zero normal flux) boundary conditions on obstacles and top/bottom walls. The Dirichlet boundary conditions are imposed on the left ($c([0,y],t)=0=c_\text{in}$) and right ($c([L,y],t)=1=c_\text{out}$) boundary.

We numerically solve the diffusion equation with the LBM to obtain the effective diffusion coefficient and concentrations. The details of the approach are given in Appendix~\ref{App:LBM}. Each LBM simulation is iterated until the maximum change of the concentration field between two consecutive iterations $\Delta c_n$ defined as:
\begin{equation}
    \Delta c_n = \max\limits_{i\>:\>\vec{r_i} \in \mathcal{D}_f} \left(\left|c_i^n - c_i^{n-1}\right|\right)
\end{equation}
is smaller than $10^{-13}$ or up to $n=10^6$ iterations. Then the effective diffusion coefficient $D$ is calculated using discrete implementation of Eq.~\eqref{eq:fick_law}, namely,
\begin{equation}\label{eq:D_fick}
    D = \frac{L}{N\left(c_\text{out}-c_\text{in}\right)} \> \sum\limits_{\vec{r}_i \in \mathcal{D}_f} \left(\nabla c\right)|_{\vec{r}_i},
\end{equation}
where $c_i^n$ denotes concentration value at node $i$ at iteration $n$. Note that we choose the LBM parameters in a way that $D_0=1$.

\begin{figure}[!ht]
    \begin{center}
    \includegraphics[width=0.8\textwidth]{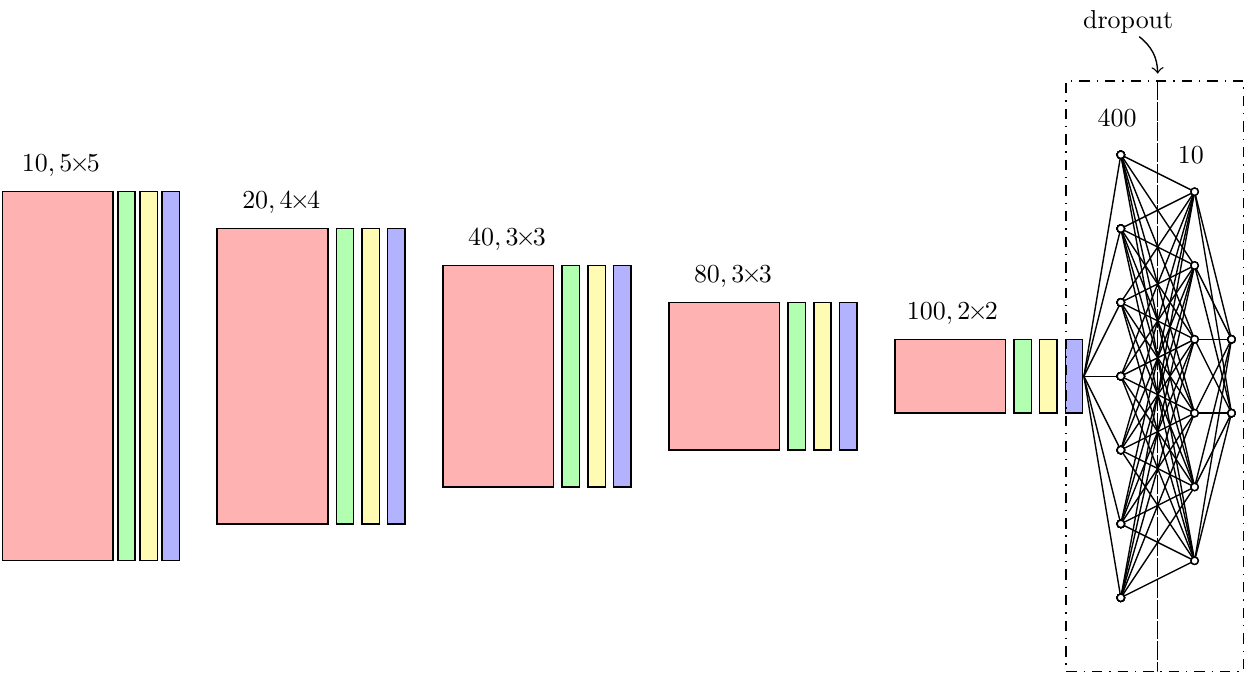}\end{center}
    \caption{ The C-Net architecture. It contains five convolutional sections. Each of them consists of the convolutional layer (red) with ReLU,  the stride one and the padding zero, batch normalization layer (green), max pooling layer with kernel $2\times2$ (violet), and dropout layer (yellow). Over each convolutional layer, there is a number and a size of kernels specified. There is a dropout layer between the last two fully connected layers with $400$ and $10$ hidden units. For the fully connected layer with 400 units, $\tanh $ is the activation function, while for the layer with $10$ units, the linear activation function is chosen. 
    \label{Fig:C-Net} }
\end{figure}

We generate the data for two tasks: predicting  $\varphi$ and $D$ from geometry and predicting the concentrations from the geometry. For both of the tasks, we produce type-A and type-B data. Each generated dataset is split into training and validation datasets.
In Figs.~\ref{fig:rect_porous_stat} and \ref{fig:voronoi_porous_stat}, we show the distributions of porosity values $\varphi$, and effective diffusion coefficients $D$, for $2017$ samples of type-A and $640$ samples of type-B.

\section{Diffusion phenomenon from deep learning}
\label{Sec:DNN}

\begin{figure}
    \begin{center}
    \includegraphics[width=0.35\textwidth]{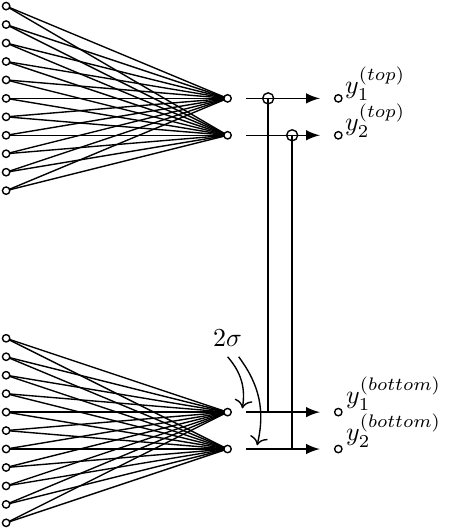}
    \caption{The self-normalized module for the  C-Net (or U-Net-Half networks). The last hidden layer of the C-Net (or U-Net-Half), outlined part of the Fig.~\ref{Fig:C-Net}, is replaced by the SN module. \label{Fig:SN-module-scheme} } 
    \end{center}
\end{figure}
\begin{figure}
    \begin{center}
    \includegraphics[width=0.99\textwidth]{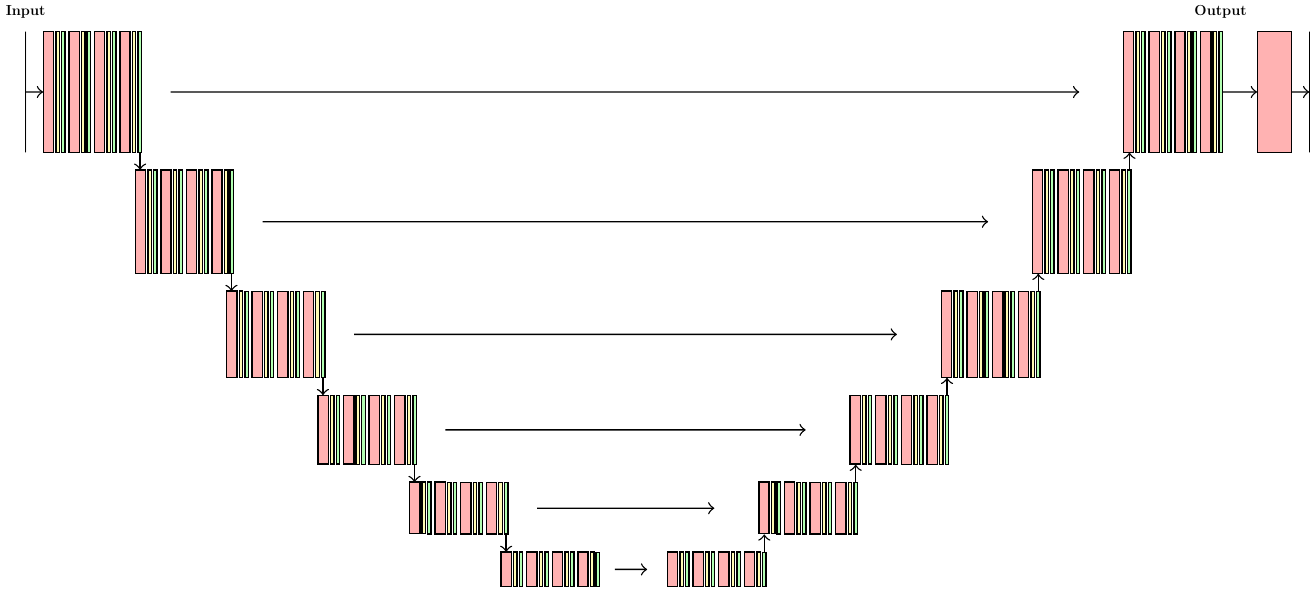}
    \end{center}
    \caption{The U-Net architecture. 
    Each downsampling container consists of ($N=4$) blocks containing a convolutional layer (with ReLU activation,  64 kernels of  $3\times3$ size),  dropout (yellow), and batch normalization layers (green). The padding in the convolutional layer is one. After each downsampling block, there is a max pooling (with kernel $2\times 2$) visualized by vertical arrow. Similarly, each upsampling container comprises four blocks of the same structure as the downsampling block. After four such blocks, there is upsampling (by two) layer (vertical arrow). Note that the upsampling container connects with the previous container's output and the corresponding block's output in the downsampling section. In the final section of the network, there is the downsampling container (without the max pooling) and convolutional layer, which reproduces a one-channel density map of the same size as the input map.\label{Fig:U-Net-Schme} }
\end{figure}

In the present paper, we exploit the deep neural network  to construct the computational system that predicts:
\begin{enumerate}
    \item[I] porosity, $\varphi$, and effective diffusion coefficient $D$, see Eq.~\ref{eq:effdiffusion};
    \item[II] the distribution of the concentrations $c(\vec{r},t)$ in the system.
\end{enumerate}
In both tasks, the network takes a picture of the system's geometry for input. In the first, the network predicts $\varphi$, $D$, and in the other, the final concentration map. 

In the following sections, we shall discuss tasks I and II separately. The  procedure for obtaining a DNN for task I can be summarized as follows: 
\begin{enumerate}
\item[(i)] Consider the samples of the given type of data; 

\item[(ii)] Each sample has assigned porosity and diffusion coefficient;

\item[(iii)] Train the DNN on training dataset, and validate on validation dataset  

\item[(iv)] Test the obtained DNN on the other data type. 
\end{enumerate}
For task II the network reconstructs the concentration field. Therefore, the data contains the pairs of figures: the input picture with the system's geometry and the output picture with the concentration field. 

The full analysis is conducted in PyTorch~\cite{NEURIPS2019_9015}.

\subsection{DNN models}

\label{Sec:Sec:DNN:models}

%
%


For task I, we consider two types of network architecture, namely:
\begin{itemize}
    \item C-Net: a network of similar structure as the one used in our previous paper~\cite{Graczyk20} designed to predict the porous medium's porosity, permeability, and (hydraulic) tortuosity.
    
    \item U-Net-Half:  it is the decoder part of the U-Net network \cite{ronneberger2015unet} exploited by us to reconstruct the maps of concentrations in the medium.
\end{itemize}

The C-Net is the convolutional neural network containing five blocks; see Fig.~\ref{Fig:C-Net}. On the top of the network are two subsequent fully connected layers with $400$ and $10$ hidden units and tanh and linear activations functions, respectively. The network takes the input picture of size $128\times128$ with geometry's configuration. Its output is a two-dimensional vector $(\varphi, D)$. Each convolutional block contains a convolutional layer (with ReLU activation function), followed by batch normalization, and dropout layers, and the max pooling (with $2\times2$ kernel) layer. After the section of the convolutional blocks, there are two fully connected layers. After the first fully connected layer, there is a dropout layer. The convolutional layers   consists of  $10$, $20$, $40$, $80$ and $100$ kernels, of size $5\times5$, $4\times4$, $3\times3$, $3\times3$, and $2\times2$, respectively. More details about the structure of the C-Net can be found in the caption of Fig.~\ref{Fig:C-Net}.

The U-Net-half is the encoder part o the U-Net (described below), followed by two fully connected layers of the same structure as in the C-Net.

Eventually, for task I, we consider two more network architectures obtained from the C-Net and the U-Net-Half, by replacing the last hidden layer with the self-normalized (SN) module. It is the structure proposed by Graczyk \textit{et al.} \cite{Graczyk2022} to correct the network's output. 
The SN mechanism is motivated by the observation that getting the deep neural network with good qualitative predictions is usually simple. But on the quantitative level, the system systematically overshoots or undershoots. The self-normalization module corrects the network response and increases accuracy in the predictions. This module is a part of the system, and its parameters are a subject of optimization too.  

The SN module is shown in Fig.~\ref{Fig:SN-module-scheme}. In practice, there are two similar-size, parallel, disconnected (from each other) fully-connected layers instead of  one fully-connected layer. The top fully-connected layer has two-dimensional output multiplied by the corresponding output of the bottom layer, namely, 
\begin{equation}
\label{Eq:SN-module-1}
(y_1^{(top)},y_2^{(top)}) \to (y_1^{(top)}y_1^{(bottom)},y_2^{(top)} y_2^{(bottom)}),
\end{equation}
where 
$(y_1^{(bottom)},y_2^{(bottom)})$ is the bottom output, where
\begin{equation}
y_i^{(bottom)} =  2 \sigma(..),
\end{equation}
$\sigma$ is the sigmoid function. With such a definition, $y_i^{(bottom)}$ plays a role of normalization factor ranging from $0$ to $2$. In principle, these factors should be around one. To enforce the network to have proper normalization, we consider the loss function with an additional penalty term, namely,
\begin{equation}
\label{Eq:SN-module-2}
L(s_i) \to L(s_i) + (1 - y_1^{(bottom)})^2 + (1- y_2^{(bottom)})^2,
\end{equation}
where $L(s_i)$ is the loss for the $s_i$ sample. For $L(...)$, we consider the mean-square error (MSE) loss in our analysis.

\begin{figure}[t]
\centering
\includegraphics[width=0.8\textwidth]{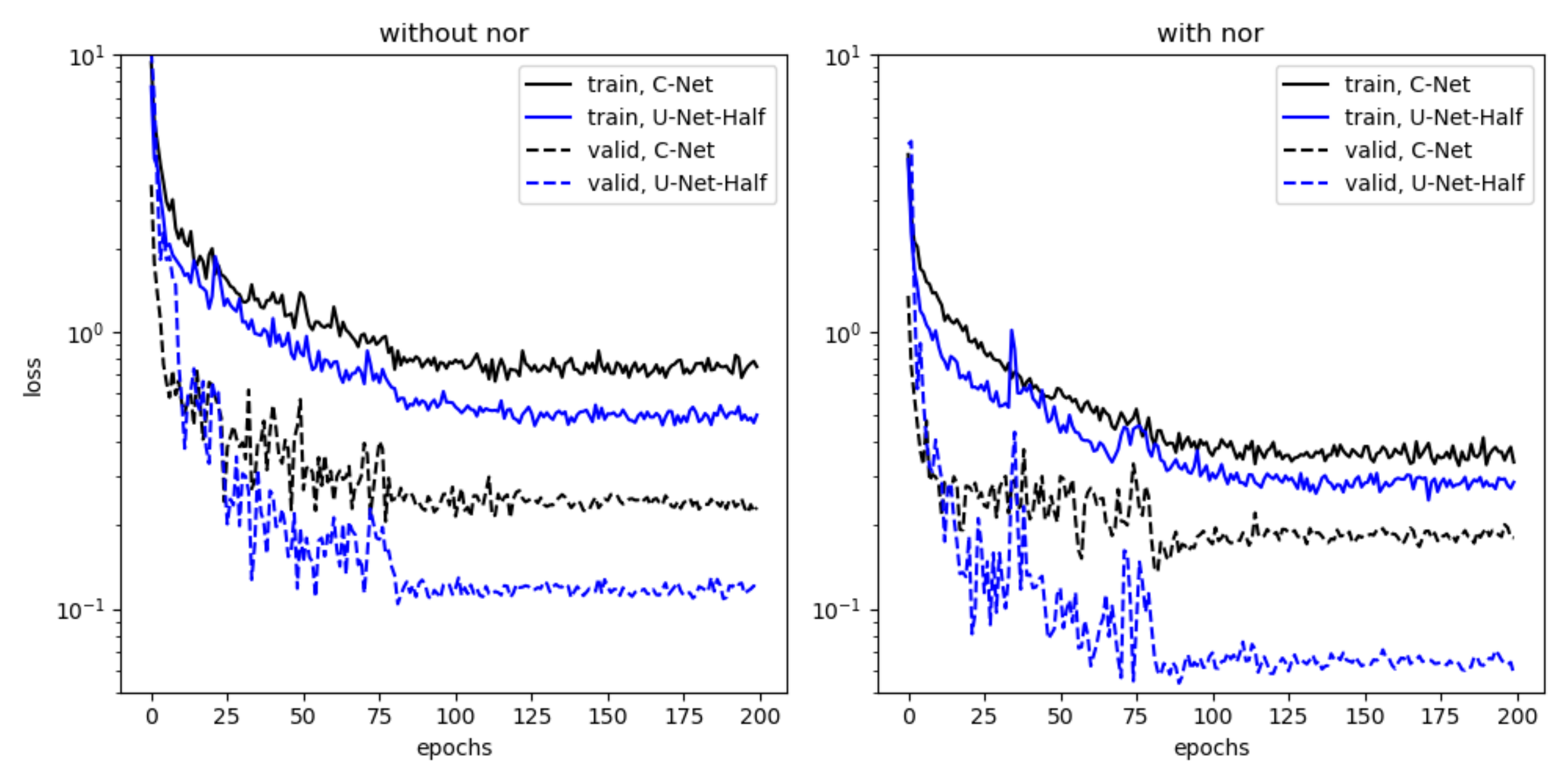}
\caption{The MSE loss during networks training. The dashed and solid curves correspond to an error on the validation and training dataset, respectively The blue and black plots correspond to the results for the U-Net-Half and the C-Net, respectively. All networks contain dropout layers with $p=0.1$. The results for the models without/with the self-normalization module are shown in the left/right figure.
\label{Fig:history_loss_scenario_one_anizo}
}
\end{figure}

For task II, we propose considering the U-Net architecture, designed to face the biomedical image segmentation problem \cite{ronneberger2015unet,Graczyk2022}. This type of network architecture was utilized, i.e., to predict the membrane's flow properties, taking its morphology as input \cite{Karmava2021c} and simulating a fluid flow in the porous materials\cite{Kamrava2021}. It has a more complicated structure than the C-Net network; see Fig.~\ref{Fig:U-Net-Schme}. In its structure, we can distinguish an encoder and a decoder part. The encoder consists of a sequence of six containers. Each of them downsamples the input by two. Each container has four blocks containing a convolutional layer (with ReLU activation), dropout, and batch normalization layers. After four blocks, the max pooling layer downsamples the input. The decoder includes six corresponding upsampling containers, and each upscales the input size by a factor of two. The upsampling container consists of four convolutional blocks (the same structure as in the encoder), followed by the upsampling layer. The encoder and decoder containers of the same output-input size are connected. Eventually, there is a block of convolutional layers (with batch normalization and dropout layers) after the decoder section. Fig.~\ref{Fig:U-Net-Schme} contains a detailed description of the U-Net structure. Note that each downsampling or upsampling container has $N=4$ convolutional blocks. In the pre-analyses, we started with $N=1$, and after several tests, we obtained an optimal value of $N=4$.

To estimate how uncertain the network predictions are, we exploit the Monte Carlo (MC) dropout technique \cite{gal2015dropout,gal2015dropout-appendix}. Graczyk \textit{et al.}\cite{Graczyk2022} adopt this method to  estimate uncertainties for network predictions in the analysis of microbiological data. In short, to get $1\sigma$ uncertainty for the network response, we run network $M=20$ times, keeping the dropout layers active. Then the model response is given by the average overall network responses, while from the standard deviation, the $1\sigma$ uncertainty is obtained.

\subsection{Porosity and diffusion coefficient from deep learning}

\label{Sec:Sec:Porosity_and_Diffusion}

Our first task is to obtain the system that predicts the porosity and diffusion coefficient based on the input picture of the system's geometry. Here we distinguish two scenarios for model development. In the first scenario, the networks are trained on type-A data, and the validation dataset is of the same kind. Then we verify how the model works on the type-B data. The second scenario refers to the models trained and validated on the type-B dataset and tested on the type-A dataset.

Note that the loss is given by the mean-square error (MSE). In the data augmentation, we exploit the property that a horizontal flip of the input picture does not change the porosity and diffusion coefficient of the system. In Fig.~\ref{Fig:history_loss_scenario_one_anizo}, we show how the value of the loss varies during the training. The plots are obtained for all discussed network models for the first scenario - training and validation datasets are of type-A. In this analysis, we set $p=0.1$ in the dropout layers. It turned out to be the most optimal value for the dropout mechanism.

\begin{figure}[!ht]
\begin{center}
\includegraphics[width=\textwidth]{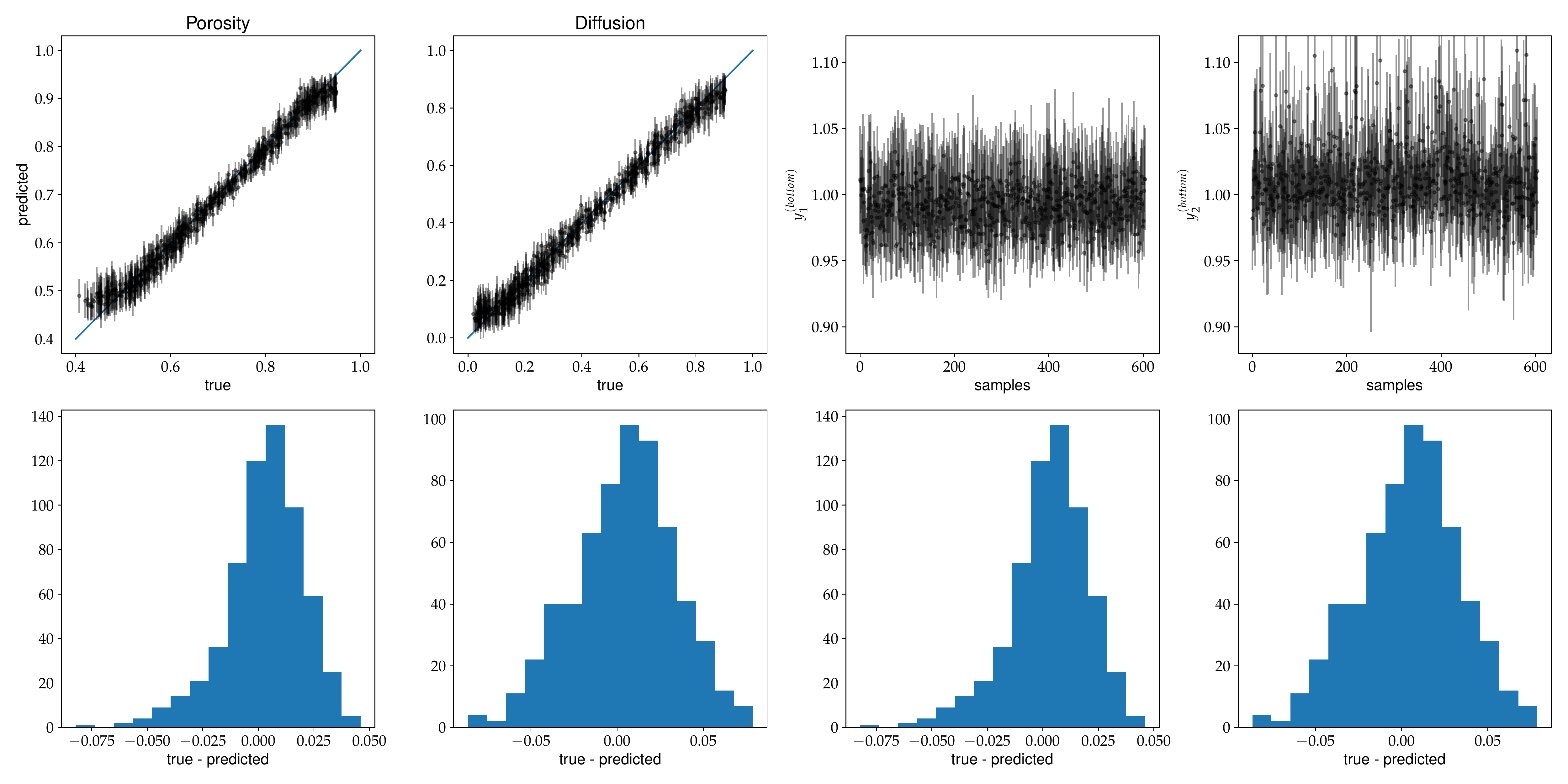}
\end{center}
    \caption{The C-Net with SN-module trained on type-A data. In the top row (the first and the second figure from the left) the predicted values of porosity and diffusion coefficient vs. true values for validation type-A data are shown. The figures third and fourth of the top row show  the normalization outputs of the SN-module, namely, $y^{bottom}_1$ and $y^{bottom}_2$, see Eq. \ref{Eq:SN-module-1}. In the bottom row the corresponding histograms of true-predicted  are given. The uncertainties are computed from the MC dropout with $p=0.1$. \label{Fig:C-Net-nor-traintest_pred_vs_true}}
\end{figure}
%
\begin{figure}[!ht]
\begin{center}
\includegraphics[width=\textwidth]{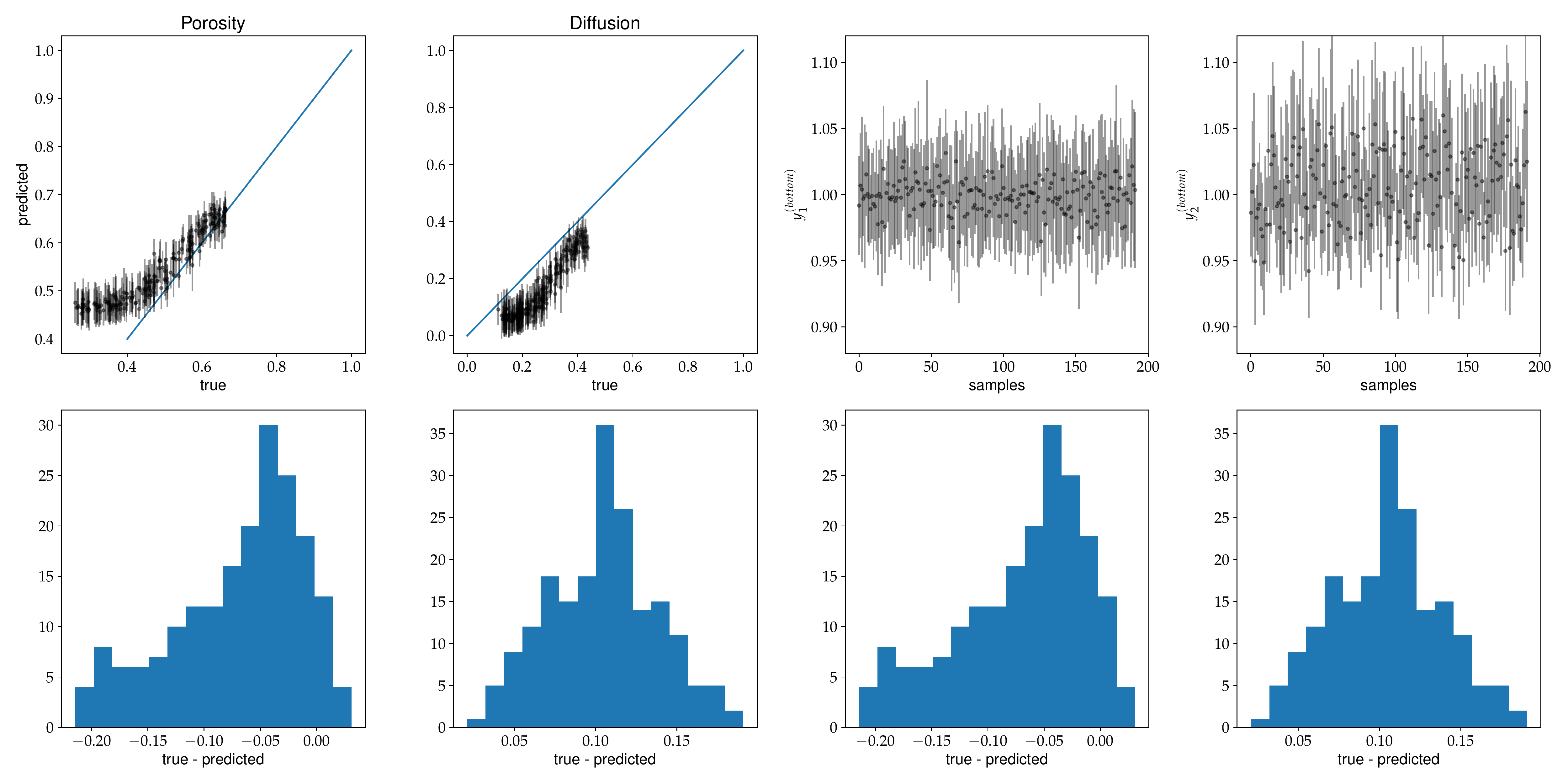}
\end{center}
\caption{Caption the same as for Fig.~\ref{Fig:C-Net-nor-traintest_pred_vs_true} but the predictions made for validation type-B data. \label{Fig:C-Net-nor-traindom-test_pred_vs_true} }
\end{figure}
\begin{figure}[!ht]
\begin{center}
\includegraphics[width=\textwidth]{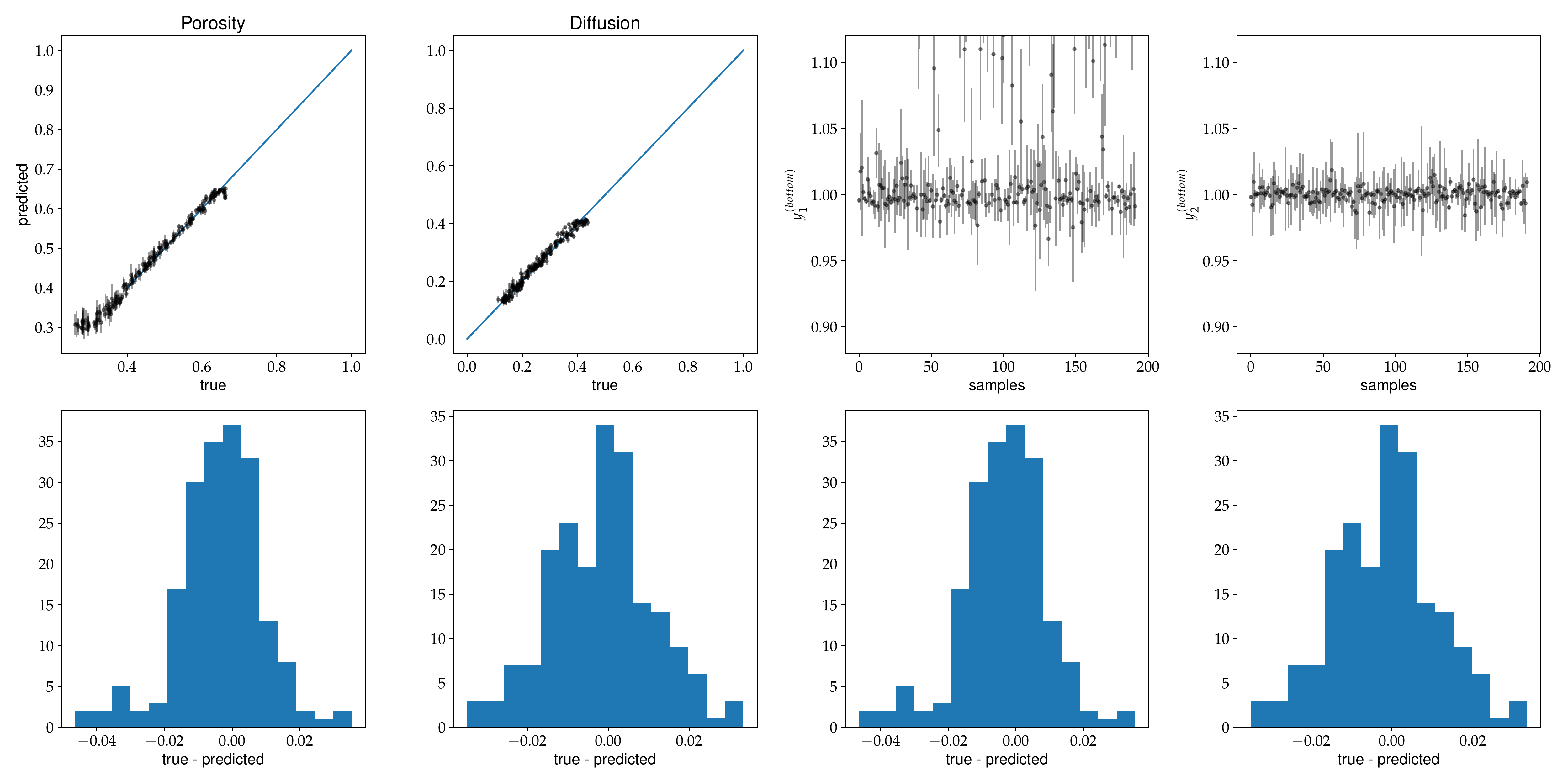}
\end{center}
\caption{Caption the same as for Fig.~\ref{Fig:C-Net-nor-traintest_pred_vs_true} but for the U-Net-Half (with SN module) trained on type-B dataset,  and the predictions made for validation type-B dataset.
\label{Fig:U-Net-Half-domtraintest_pred_vs_true}}
\end{figure}

The detailed analysis of  Fig.~\ref{Fig:history_loss_scenario_one_anizo} leads to the conclusion that adding the self-normalization module improves the performance of the networks. Indeed the errors computed on the validation dataset for the models with SN-module are systematically lower than those obtained for the vanilla networks. 
On the other hand, the errors computed for the C-Net  networks are systematically larger than for the U-Net-half models, which can be interpreted 
that the U-Net-Half works better than C-Net.
To study this problem more carefully, we introduce
\begin{equation}
\overline\chi^2 = \frac{1}{N}\sum_{i=1}^N\left[ \frac{( \varphi_{pred}^i - \varphi_{true}^i)^2}{\Delta\varphi^{i\,2}} + \frac{( D_{pred}^i - D_{true}^i)^2}{\Delta D^{i\,2}} \right],   
\end{equation}  
where $\varphi^i_{pred/true}$ and $D^{i}_{pred/true}$ denote the predicted/true values for porosity and diffusion coefficient for the $i$-th sample, respectively. The $\overline\chi^2$  includes information about the uncertainties, $\Delta\varphi$, and $\Delta D$, obtained from the MC dropout. 

For the C-Net and the U-Net-Half (both with SN module) models trained on type-A dataset, we obtained $\overline\chi^2 = 1058.4$ and $\overline\chi^2 = 1465.4$ (computed on the validation dataset type-A). In this case,  the U-Net-half tends to overfit the data more than the C-Net.
Unsurprisingly, the C-Net network is defined by a much lower number of parameters than U-Net-Half. Indeed, the size of the first network is about $300$ kB, while the other is about $3.3$ MB. Accordingly, the C-Net with SN-module is a better model for predicting porosity and diffusion coefficient for type-A dataset.
However, when the models are trained on the type-B dataset, the $\overline\chi^2$, computed on the validation dataset, takes the values: $342.7$ and $237.0$ for the C-Net and the U-Net-Half (both with SN module), respectively. Hence, for type-B data, the U-Net-Half is more certain. Moreover, the models trained on the type-B data sets are more accurate in the predictions than models trained on type-A. This property is summarized in Table~\ref{Table}, where we give the MSE errors for both analyzed scenarios.
\begin{table}
\centering{
\begin{tabular}{c||c|c}
     & C-Net with SN & U-Net-Half with SN\\
     \hline\hline
   MSE (type-A data)   & $0.73$ &  $0.72$ \\
   MSE (type-B data)   & $0.33$ &  $0.29$
\end{tabular}
\caption{The MSE error computed on validation datasets, for the models run in the Monte Carlo dropout framework. \label{Table}}
}
\end{table}

\begin{figure}[!ht]
\centering
\includegraphics[width=.7\textwidth]{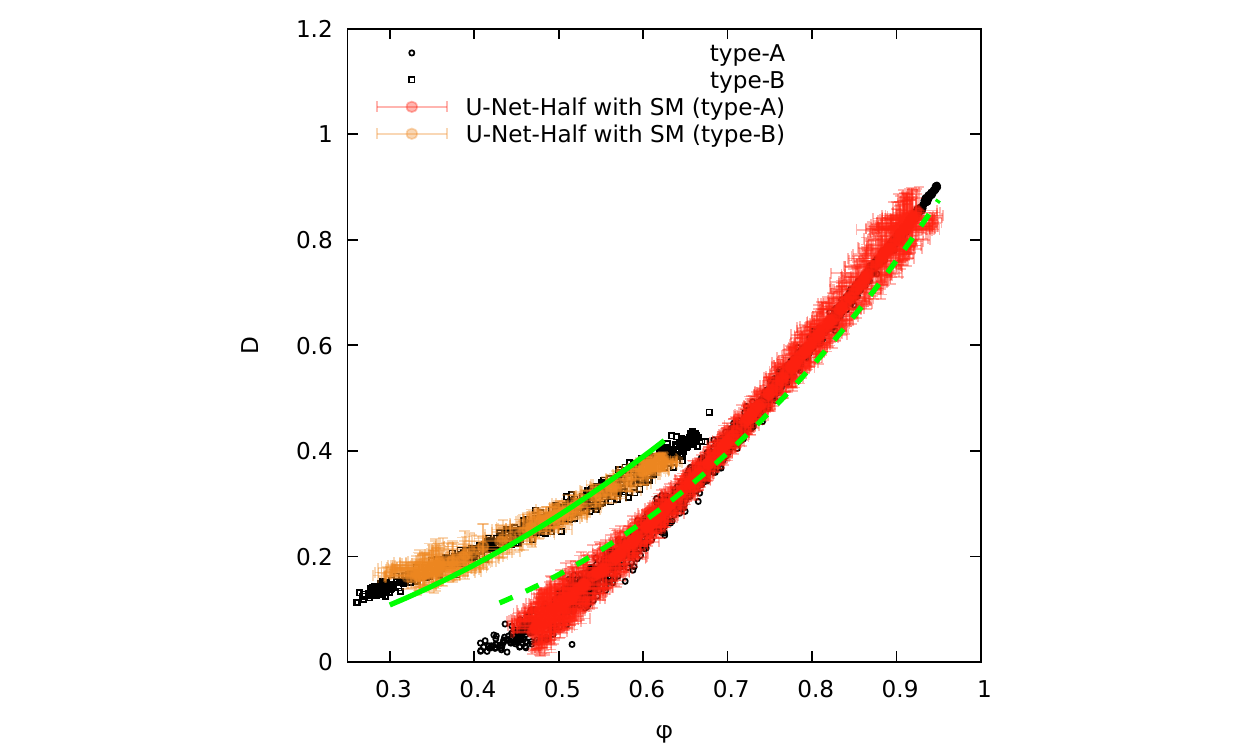}
\caption{Effective diffusion against true porosity.  Open symbols denote the data from LBM solution,  filled circles with errorbars denote predictions of the two U-Net-Half (with SN module) networks trained on type-A or type-B datastes. The predictions are made for the validation data sets of type-A or type-B, respectively.
Lines (green in color) denote the fits of the effective diffusion as a function of porosity based on Archie law, see Eq.~\ref{Eq:Archie}. }
\label{fig:D_vs_porosity}
\end{figure}

Let us discuss the quality of our fits in a more detailed way.
In Fig.~\ref{Fig:C-Net-nor-traintest_pred_vs_true}, we present the predictions for the validation dataset of type A  of the  C-Net trained on the same type of data.
A good match between the predictions and true values of $(\varphi,D)$ is observed. It is not the case, when one makes predictions for type-B dataset, see  Fig.~\ref{Fig:C-Net-nor-traindom-test_pred_vs_true}. 
The model overestimates the porosity and underestimates the diffusion coefficient in this case. However, the network's predictions are consistent within the two-sigma level for the samples with $\varphi > 0.4$. Note that type-A data does not contain samples with $\varphi < 0.4$. Indeed, it is challenging to generate type-A data with porosity smaller than  $0.4$ as it is close to the percolation threshold of discrete overlapping quads, as shown in Koza \textit{et al.}~\cite{Koza2014}. Moreover, it is so unlikely that the algorithm generating type-A samples will produce samples of type-B data. Hence, it is no surprise that the network, trained on type-A data, does not work for type-B data, especially in the porosity range not accessible by samples of type-A.

\begin{figure}
    \centering
    \includegraphics[width=0.496\textwidth]{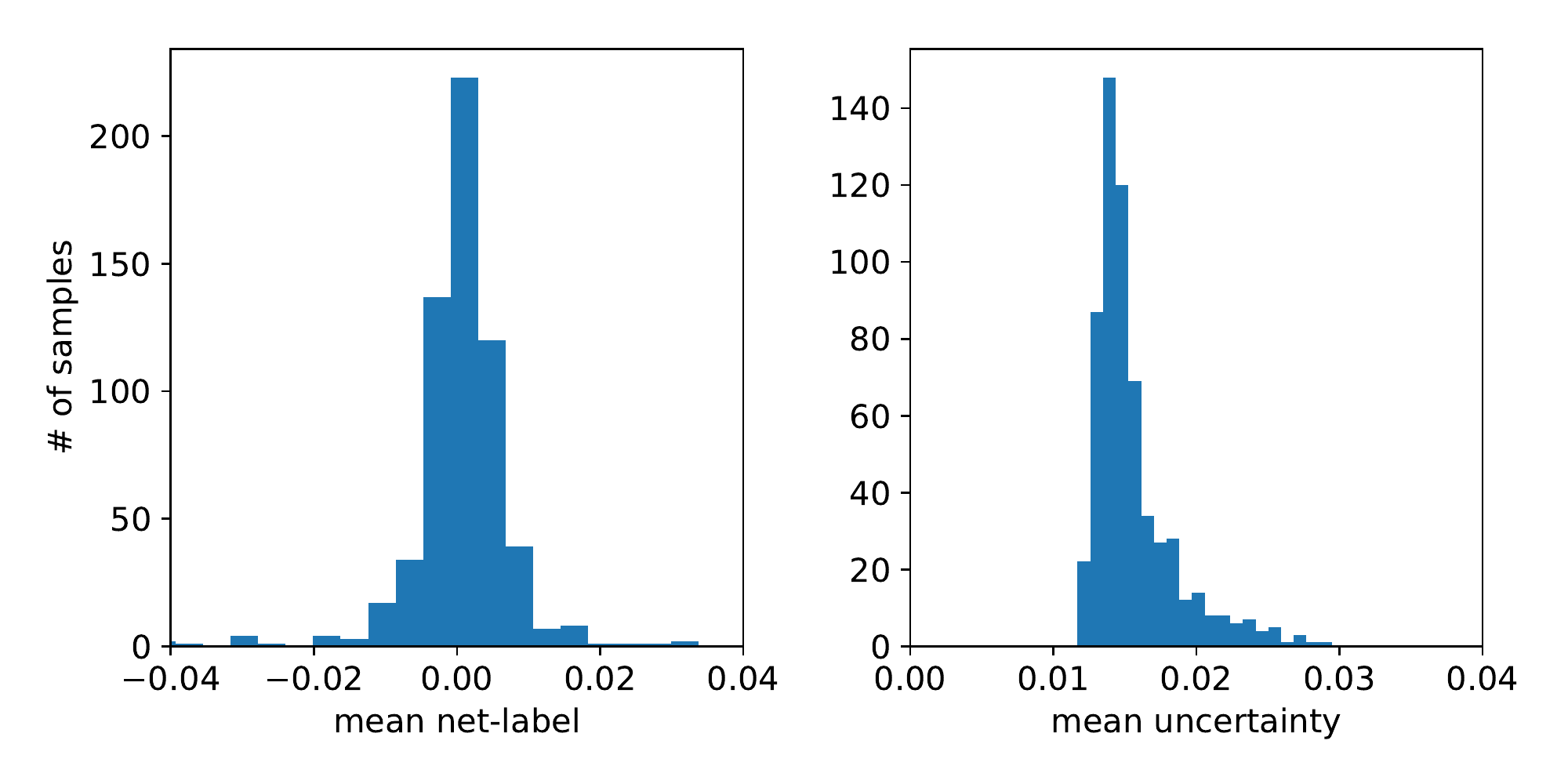}
\includegraphics[width=0.496\textwidth]{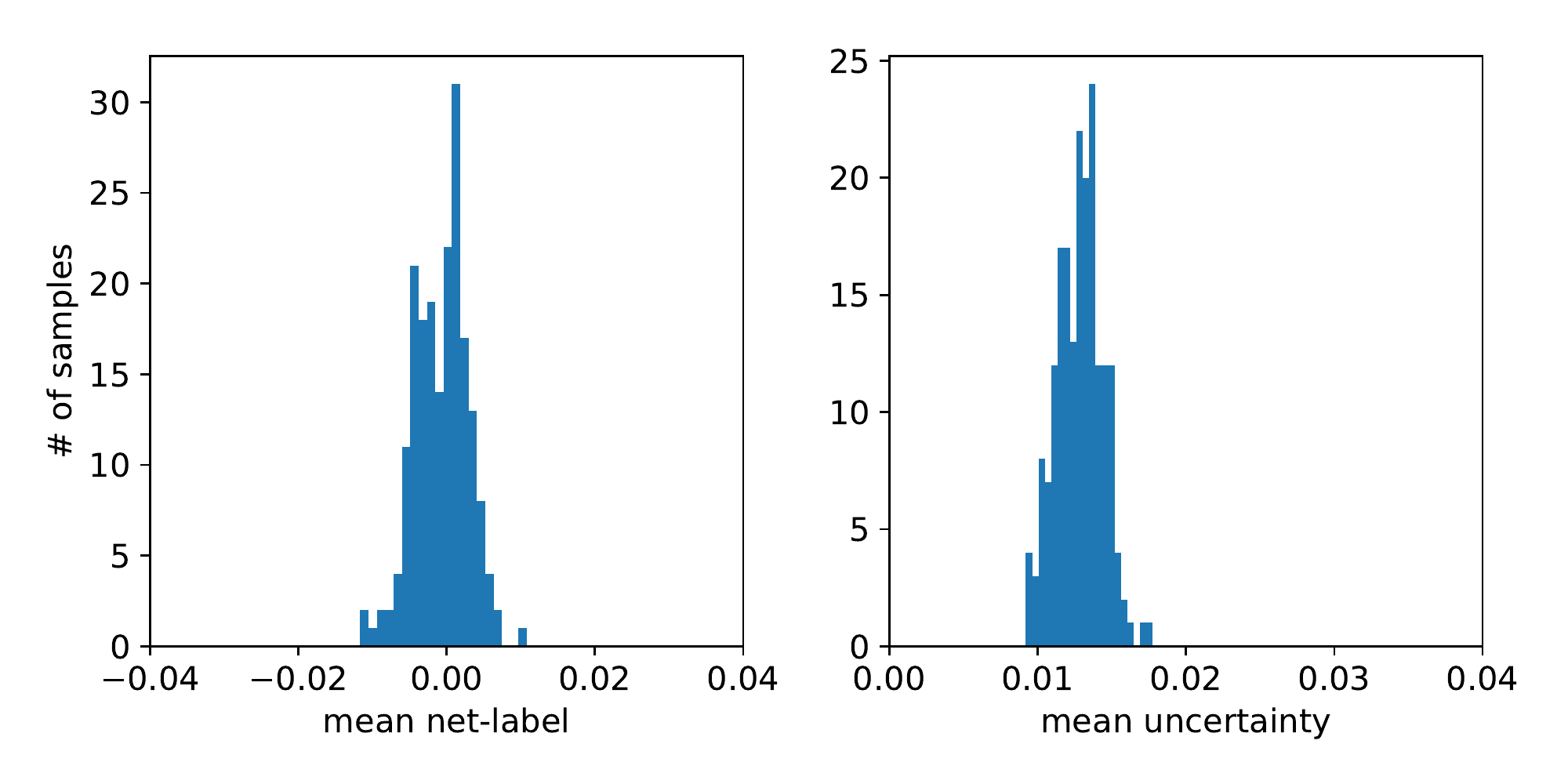}
    \caption{The summary of the predictions of concentrations by the U-Net model, trained on type-A data, for the validation datasets of type-A (two figures on the right), and type-B (two figures on the left).}
    \label{Fig:U-Net_tests_trained_on_ani}
\end{figure}
\begin{figure}
    \centering
\includegraphics[width=0.496\textwidth]{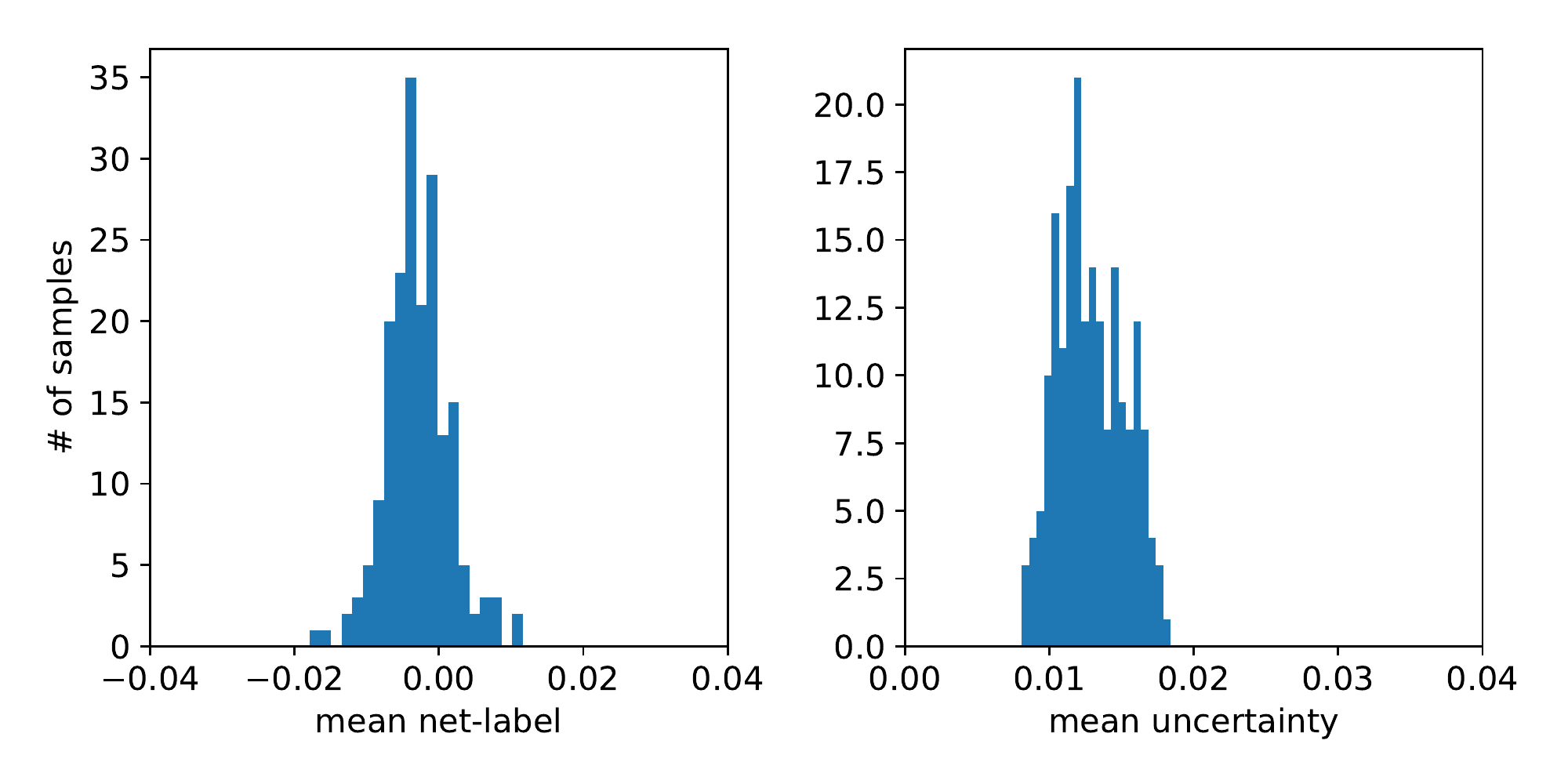}
\includegraphics[width=0.496\textwidth]{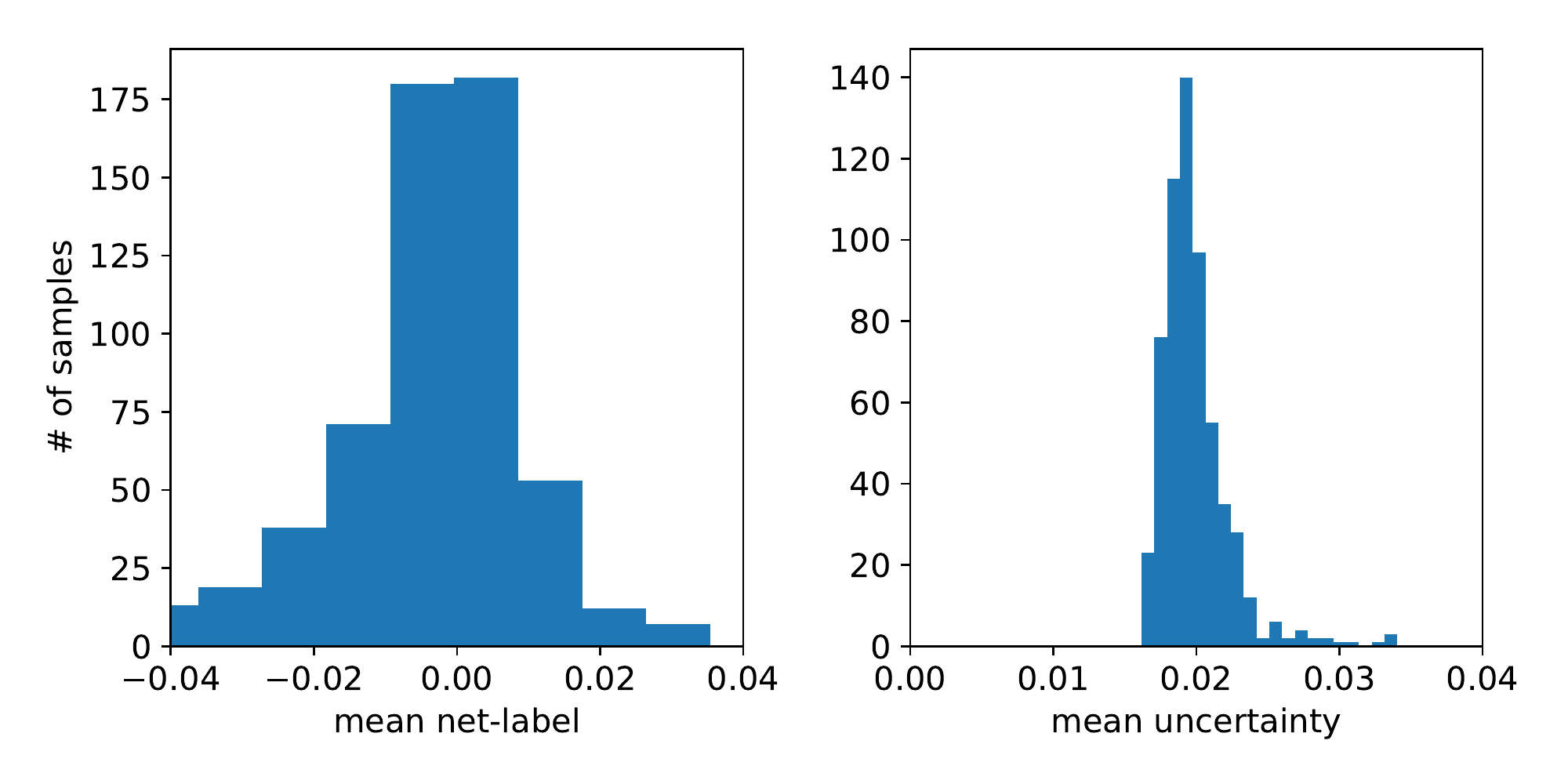}
    \caption{The summary of the predictions of concentrations by the U-Net model, trained on type-B data, for the validation datasets of type-B (two figures on the right), and type-A (two figures on the left).}
    \label{Fig:U-Net_tests_trained_on_dom}
\end{figure}

\begin{figure}[!ht]
\centering
\includegraphics[width=0.9\textwidth]{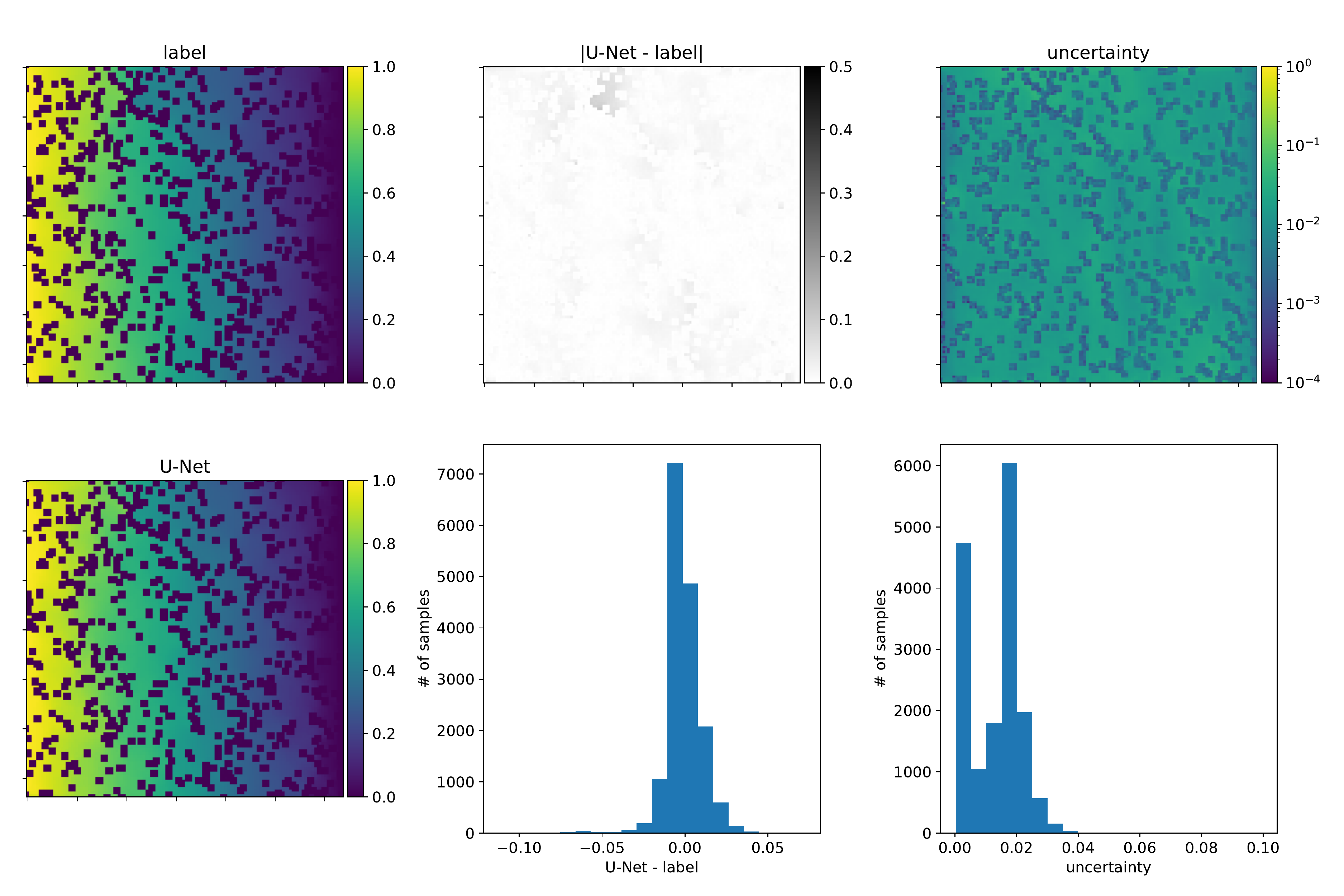}
\caption{ 
Predictions of the concentrations of the U-Net, trained on type-A data, for type-A input sample. Top row, from left to right: the input figure, the differences between the input and the network's response and a map of uncertainties for each node. Bottom row, from left to right: the network prediction, 
histogram of differences between input and network's outcome
histogram of uncertainties at each node. The direction of transport is from the left to the right. Note that the smallest errors are obtained for the nodes where obstacles are placed, while in the pore space, the uncertainties are more considerable but uniformly distributed.}
\label{Fig:maps_trainedonani_for_ani} 
\end{figure}
\begin{figure}[!ht]
\centering
\includegraphics[width=0.9\textwidth]{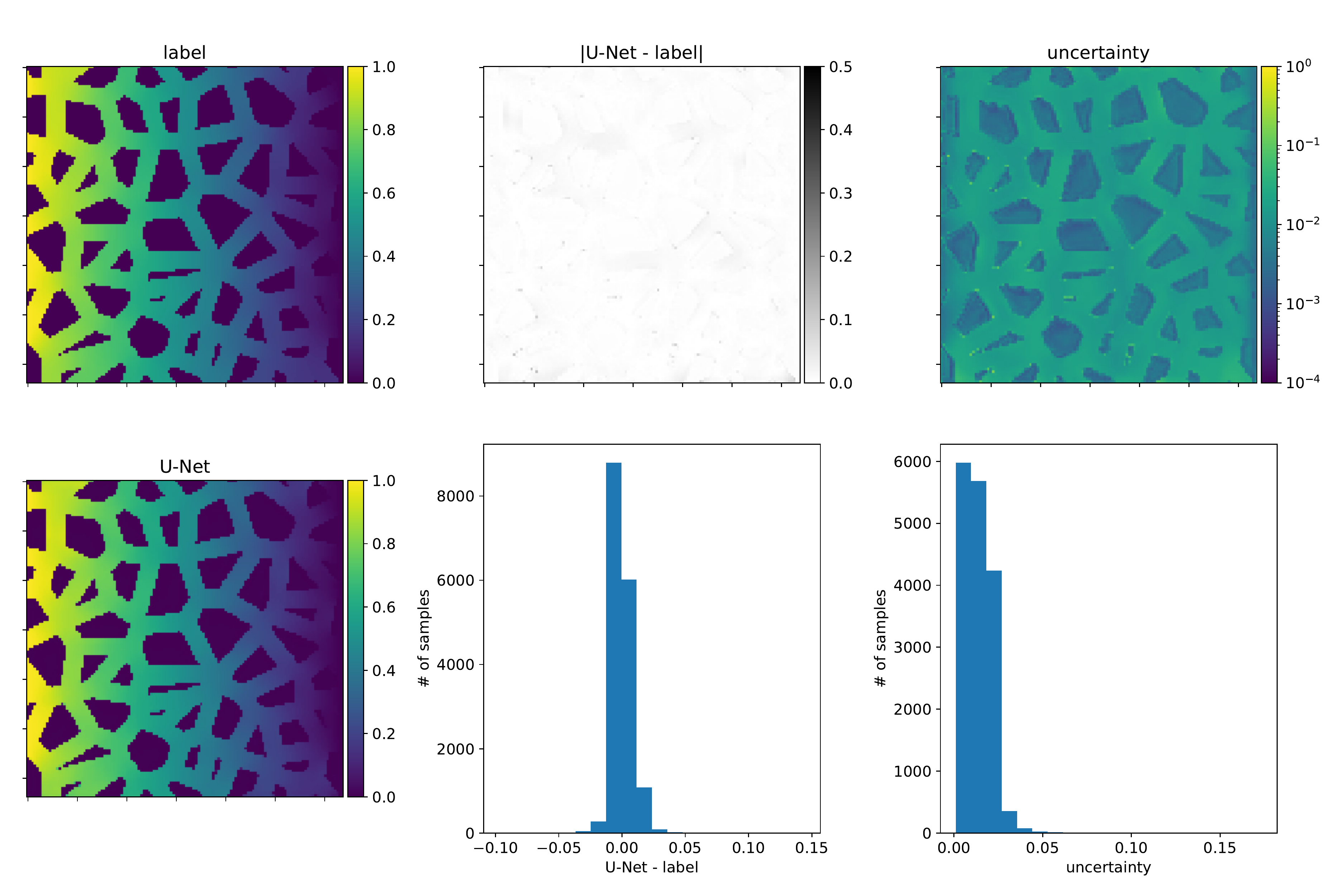}
\caption{The same as  Fig.~\ref{Fig:maps_trainedonani_for_ani} but for the model trained on type-A data, and the  predictions made for type-B input sample. \label{Fig:maps_trainedonani_for_dom} }
\end{figure}

In the second scenario, the models are trained on type-B data. As stated above, U-Net-Half works better than the C-Net architectures in this case.   
Fig.~\ref{Fig:U-Net-Half-domtraintest_pred_vs_true} presents the predictions of the U-Net-Half (with SN module)  
for the type-B validation dataset. The network reproduces the data well, even in the low porosity range. However, the model does not work for type-A data samples (not shown in the paper). 

\begin{figure}[!ht]
\centering
\includegraphics[width=0.9\textwidth]{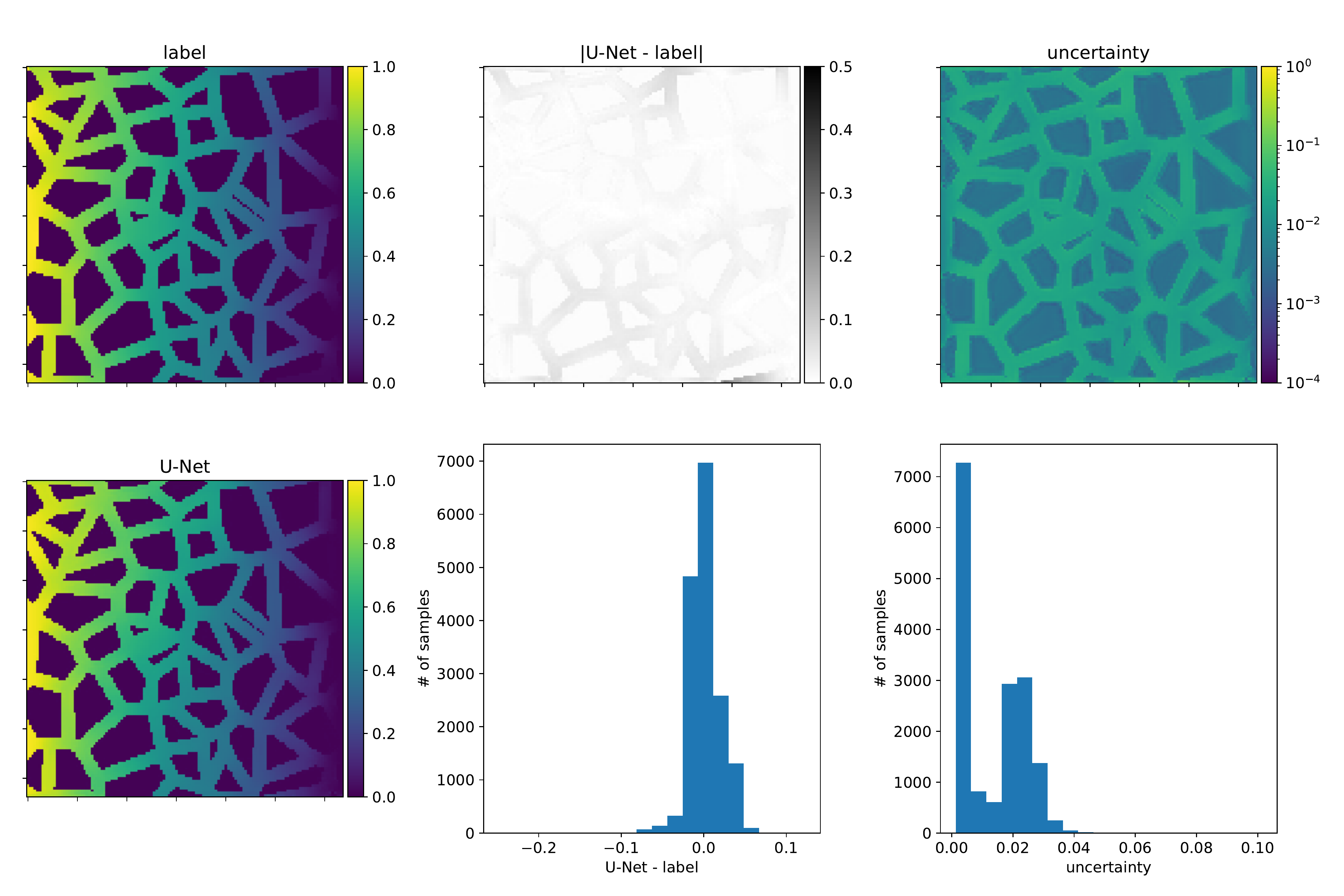}
\caption{ The  same as for Fig.~\ref{Fig:maps_trainedonani_for_ani} but for the model trained on type-B data, and the  predictions made for type-B input sample. \label{Fig:maps_trainedondom_for_dom}  }
\end{figure}
\begin{figure}[!ht]
\centering
\includegraphics[width=0.9\textwidth]{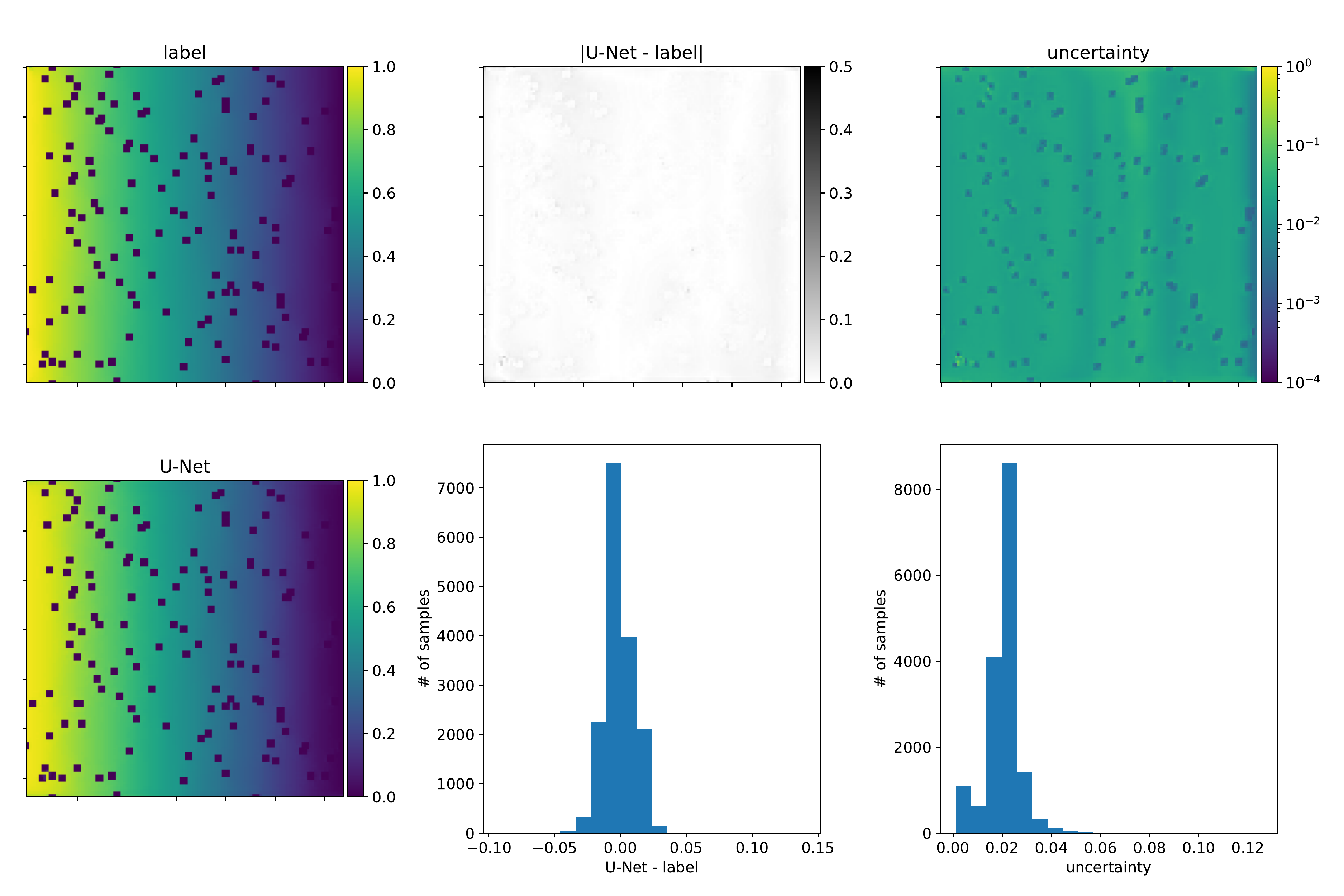}
\caption{ The  same as for Fig.~\ref{Fig:maps_trainedonani_for_ani} but for the model trained on type-B data, and the  predictions made for type-A input sample. \label{Fig:maps_trainedondom_for_ani} }
\end{figure}

As the final cross-check, we present a plot of a diffusion coefficient as a function of true porosity, see Fig.~\ref{fig:D_vs_porosity}. The data obtained with the LBM  and predictions of the two U-Net-Half (with SN module) models trained on type-A and type-B data agree reasonably. 

The data show a correlation between the diffusion coefficient and porosity. In Fig.~\ref{fig:D_vs_porosity}, we plot the fits of $D(\phi)$ as well. To fit the data, we assumed that
$D$ is the reciprocal of tortuosity squared as defined in\cite{Boudreau1996}. Then, we used the Archie's law to express tortuosity, $\lambda$: 
\begin{equation}
\label{Eq:Archie}
\lambda^2=\varphi^{1-n},
\end{equation}
where $n$ is the Archie's exponent\cite{Boudreau1996}. We found $n=3.6$ for type-A data (dashed line) and $n=2.8$ for type-B data (solid line), which corresponds well to $n=2.14$ reported by Boudreau \cite{Boudreau1996} for fine-grained sediment.


\subsection{Reconstruction of the concentrations by U-Net}

\label{Sec:Sec:DNN:Concentration}

This subsection summarizes the results obtained for task II. Our goal was to bring the network, which takes as an input the figure with system geometry configuration, and as an output, it gives the picture with geometry and distribution of concentration. The training data consists of pairs of pictures: input geometry and concentration map. 

Similarly, as in task I, we consider two scenarios. First, we train and validate the network on type-A data. Then, we test the model on the type-B dataset. Second, we train and validate the network on type-B data and test the model on the type-A  dataset. We adopt the MC dropout technique (with $p=0.1$) to estimate uncertainties as in the previous investigations.

To assess the obtained results for every picture sample, we compute the difference between the input and output pictures and the mean over all nodes. Note that for each site, we obtain $1\sigma$ uncertainty, and to estimate the average uncertainty for the output picture, we compute the mean over all nodes, see Figs.~\ref{Fig:U-Net_tests_trained_on_ani}~and~\ref{Fig:U-Net_tests_trained_on_dom}. 

As can be seen, Fig.~\ref{Fig:U-Net_tests_trained_on_ani}, the U-Net, trained on type-A data, predicts the concentration field well.
Unlike task I, the model works well for data type B too. Indeed, in task I, the model trained on type-A data could not predict the porosity and diffusion coefficient for samples with $\varphi<0.4$,  which is reached only by type-B samples. Moreover, the U-Net trained on type-B data samples accurately predicts the concentration field for the type-B and type-A data samples, see Fig.~\ref{Fig:U-Net_tests_trained_on_dom}.

We finish the discussion of the numerical results with the presentation of the U-Nets' performance on chosen samples. Figs.~\ref{Fig:maps_trainedonani_for_ani}~and~\ref{Fig:maps_trainedonani_for_dom} present results for the network trained on type-A data, its predictions of the concentration field for type-A and type-B data samples, respectively. Figs.~\ref{Fig:maps_trainedondom_for_dom}~and~\ref{Fig:maps_trainedondom_for_ani} show the analogical results as before but for the network trained on type-B data, and its prediction for type-B and type-A samples, respectively. Each figure contains two rows of plots. The first row consists of a map of the true concentration field, a map of absolute error between true and predicted, and a map of uncertainties. The second row contains a map of the predicted concentration field and
histograms of absolute error and uncertainties. One sees that the network prediction error is small for all four cases compared to the maximum concentration value in the LBM solution. The qualitative comparison of mapping obstacles' shapes and positions shows that the networks handle this task well. The error (true-predicted) distributions in each figure are concentrated around the value of zero. For most of the samples, the uncertainty for the network predictions in the void space is more prominent than in the obstacle positions; see Figs.~\ref{Fig:maps_trainedonani_for_ani}-\ref{Fig:maps_trainedondom_for_ani}.

%


\section{Summary}
\label{Sec:Summary}
We have developed the DL models to predict the basic properties of diffusion in the porous media. We considered two different types of CNNs. Both take as the input the picture of the system's geometry, but one predicts the porosity and diffusion coefficient, while the other reconstructs the systems' geometry and concentration map.

For the first task, we discussed two variants of the CNN: the C-Net and the U-Net-Half. Both are accompanied by the self-normalized module proposed by Graczyk \textit{et al.}\cite{Graczyk2022}. We show that the models with the SN module have better performance than those without the module. For the second task, we proposed to consider the U-Net architecture. The model  perfectly reconstructs the concentration field for all types of data.

We have considered two types of data: one mimics the sand packings, and the other mimics the systems derived from the extracellular space of biological tissues. One of our goals was to verify whether the network, trained on one type of data, can be used for analysis in the case of the other and conversely. The models that predict the porosity and diffusion coefficient work well only on the data type they were trained on. The model's porosity and diffusion coefficient predictions are within $1\sigma$ confidence level bounds.

In contrast to the task I, the models developed for task II (reconstruction of the concentration map) work well for the data type that was not included in the training process. The best accuracy in the reconstruction of the concentration maps is achieved for type-B data, namely, systems derived from the extracellular space of biological
tissues. In this case, the mean error is smaller than $1\%$ while  the mean map reconstruction uncertainty is slightly larger.

\section*{Availability of Data and Materials}
The dataset used during the current study is available on request, please contact Krzysztof Graczyk.

\section*{Acknowledgments}
D.S. and M.M supported by National Science Centre.
Funded by National Science Centre, Poland under the OPUS call in the Weave programme 2021/43/I/ST3/00228.

K.M.G. has been partly supported by the program ''Excellence initiative - research university'' for the years 2020-2026 for the University of  Wroc\l{}aw.

The publication was partially financed by the Initiative Excellence–Research University program for University of Wroclaw.

This research was funded in whole or in part by National Science Centre (2021/43/I/ST3/00228).

For the purpose of Open Access, the author has applied a CC-BY public copyright licence to any Author Accepted Manuscript (AAM) version arising from this submission.

\appendix

\section{Lattice Boltzmann method}

\label{App:LBM}

Lattice Boltzmann Method\cite{Krueger2016,Succi2018} is a numerical tool for solving discrete Boltzmann transport equation:
\begin{equation}
    \frac{\partial f_k}{\partial t} + \vec{\xi}_k \cdot \nabla f_k = \Omega(f_k^\text{eq},f_k); \quad f_k=f_k(\vec{r},t)
\end{equation}
where $f_k$ is the distribution function discretized along $k$-th population velocity and $\Omega$ is collision operator. Depending on the set of discretized velocities $\vec{\xi}_k$ and the used equilibrium functions $f_k^\text{eq}$, certain moments of the equation's solution are also the solution of macroscopic transport equations. For example, in two dimensional space ($\vec{r}\in \mathbb{R}^2$) with the set of nine discrete velocities:
\begin{equation}\label{eq:LBM_vel_disc}
    \begin{array}{lll}
        \bsym{\xi}_0 = [0,0]^T & \bsym{\xi}_1 = [1,0]^T& \bsym{\xi}_5 = [1,1]^T \\
        & \bsym{\xi}_2 = [0,1]^T & \bsym{\xi}_6 = [-1,1]^T \\
        & \bsym{\xi}_3 = [-1,0]^T & \bsym{\xi}_7 = [-1,-1]^T \\
        & \bsym{\xi}_4 = [0,-1]^T & \bsym{\xi}_8 = [1,-1]^T 
    \end{array}
\end{equation}
and equilibrium distribution function:
\begin{equation}\label{eq:lattice_weights}
    f_k^\text{eq} = \omega_k c, \quad \text{where} \quad
    \omega_0 = 4/9, \quad
    \omega_{1,2,3,4} = 1/9, \quad
    \omega_{5,6,7,8} = 1/36,
\end{equation}
one obtains the concentration field of diffusing molecules (solution of Eq. \eqref{eq:diffusion}) as the zeroth moment of the distribution function:
\begin{equation}
    c(\vec{r},t) = \sum\limits_{k=0}^8 f_k(\vec{r},t)
\end{equation}

In our analysis we use the set of discrete velocities from Eq.~\eqref{eq:LBM_vel_disc}, equilibrium distributions given in Eq.~\eqref{eq:lattice_weights} and the collision operator introduced by Bhatnagar, Gross and Krook \cite{Bhatnagar1954} (BGK collision):
\begin{equation}
    \Omega_\text{BGK}(f_k^\text{eq},f_k) = \frac{1}{\tau}(f_k^\text{eq}-f_k),
\end{equation}
where $\tau$ is relaxation time of a distribution function to its equilibrium. The bulk diffusion coefficient of LBM model equals then
\begin{equation}
    D_{0,\text{LBM}} = \frac{1}{3}\left(\tau-\frac{1}{2}\right).
\end{equation}
We use above quantity to calculate macroscopic diffusion coefficient of the solved diffusion equation:
\begin{equation}
    D_0 = D_{0,\text{LBM}} \frac{\delta x^2}{\delta t},
\end{equation}
where $\delta x \> \text{[m]}$ and $\delta t \> \text{[s]}$ are LBM physical space and time discretization parameters, respectively.

We solve the discrete Boltzmann equation with lagrangian transport, in which post-collision populations $f_k^*$ are streamed from departure nodes $\vec{r}-\vec{\xi}_k\delta x$ to arrival nodes $\vec{r}$:
\begin{equation}\label{eq:LBM_algorithm}
    \underbrace{f_k(\vec{r},t+1) = f_k^*(\vec{r}-\vec{\xi}_k,t)}_\mathrm{streaming} = \underbrace{f_k(\vec{r}-\vec{\xi}_k,t) + \frac{1}{\tau}\left(f_k^\text{eq}(\vec{r}-\vec{\xi}_k,t) - f_k(\vec{r}-\vec{\xi}_k,t)\right)}_\mathrm{collision}.
\end{equation}

Eventually, we assume reflecting boundary conditions on obstacles surfaces and on top and bottom boundary. It was realised with bounceback scheme:
\begin{equation}
    f_k(\vec{r},t+1) = f_{k'}^*(\vec{r},t) \quad \text{when} \quad \vec{r}+\vec{\xi}_{k'} \in \mathcal{D}_s
\end{equation}
\noindent where $k'$ denotes lattice vector opposite to $k$, i.e. $\vec{\xi}_{k'} = -\vec{\xi}_k$. On left and right boundaries (inlet and outlet, respectively) constant concentrations were assumed (Dirichlet boundary conditions) equal to $c([0,y],t)=c_\text{in}=1$ and $c([L,y],t)=c_\text{out}=0$ which are realized by setting the unknown populations to:
\begin{equation}\label{eq:dirichlet_bc}
    f_k = \left(c_B - \sum\limits_{i \in k_\mathcal{D}}f_i\right) \frac{\omega_k}{\sum\limits_{i \in k_{\mathcal{D}'}} \omega_i} \quad \text{for} \quad \vec{r}+\vec{\xi}_{k'} \in \mathcal{D}',
\end{equation}
where $c_B$ denotes the prescribed concentration at the boundary (0 or 1) and summation over $k_\mathcal{D}$ and $k_{\mathcal{D}'}$ denotes summing over lattice neighbors laying inside and outside of domain $\mathcal{D}$, respectively. Thus, Eq.~\eqref{eq:dirichlet_bc} amounts to distributing the concentration from outside of $\mathcal{D}$ according to lattice weights $\omega_k$ of the nodes laying outside of $\mathcal{D}$.

\bibliography{diffusion}

\begin{thebibliography}{10}
\urlstyle{rm}
\expandafter\ifx\csname url\endcsname\relax
  \def\url#1{\texttt{#1}}\fi
\expandafter\ifx\csname urlprefix\endcsname\relax\def\urlprefix{URL }\fi
\expandafter\ifx\csname doiprefix\endcsname\relax\def\doiprefix{DOI: }\fi
\providecommand{\bibinfo}[2]{#2}
\providecommand{\eprint}[2][]{\url{#2}}

\bibitem{Bell61}
\bibinfo{author}{Bell, J.} \& \bibinfo{author}{Grosberg, P.}
\newblock \bibinfo{journal}{\bibinfo{title}{Diffusion through porous
  materials}}.
\newblock {\emph{\JournalTitle{Nature}}} \textbf{\bibinfo{volume}{189}},
  \bibinfo{pages}{980--981} (\bibinfo{year}{1961}).

\bibitem{Shen2007}
\bibinfo{author}{Shen, L.} \& \bibinfo{author}{Chen, Z.}
\newblock \bibinfo{journal}{\bibinfo{title}{Critical review of the impact of
  tortuosity on diffusion}}.
\newblock {\emph{\JournalTitle{Chemical Engineering Science}}}
  \textbf{\bibinfo{volume}{62}}, \bibinfo{pages}{3748--3755}
  (\bibinfo{year}{2007}).

\bibitem{Kuhn22}
\bibinfo{author}{Kuhn, T.} \emph{et~al.}
\newblock \bibinfo{journal}{\bibinfo{title}{Single-molecule tracking of nodal
  and lefty in live zebrafish embryos supports hindered diffusion model}}.
\newblock {\emph{\JournalTitle{Nature communications}}}
  \textbf{\bibinfo{volume}{13}}, \bibinfo{pages}{1--15} (\bibinfo{year}{2022}).

\bibitem{Muñoz-Gil2021}
\bibinfo{author}{Mu{\~{n}}oz-Gil, G.} \emph{et~al.}
\newblock \bibinfo{journal}{\bibinfo{title}{Objective comparison of methods to
  decode anomalous diffusion}}.
\newblock {\emph{\JournalTitle{Nature Communications}}}
  \textbf{\bibinfo{volume}{12}}, \bibinfo{pages}{6253},
  \doiprefix\url{10.1038/s41467-021-26320-w} (\bibinfo{year}{2021}).

\bibitem{Sykova2008}
\bibinfo{author}{Sykov{\'a}, E.} \& \bibinfo{author}{Nicholson, C.}
\newblock \bibinfo{journal}{\bibinfo{title}{Diffusion in brain extracellular
  space}}.
\newblock {\emph{\JournalTitle{Physiological reviews}}}
  \textbf{\bibinfo{volume}{88}}, \bibinfo{pages}{1277--1340}
  (\bibinfo{year}{2008}).

\bibitem{Nicholson2001}
\bibinfo{author}{Nicholson, C.}
\newblock \bibinfo{journal}{\bibinfo{title}{Diffusion and related transport
  mechanisms in brain tissue}}.
\newblock {\emph{\JournalTitle{Reports on progress in Physics}}}
  \textbf{\bibinfo{volume}{64}}, \bibinfo{pages}{815} (\bibinfo{year}{2001}).

\bibitem{Postnikov2022}
\bibinfo{author}{Postnikov, E.~B.}, \bibinfo{author}{Lavrova, A.~I.} \&
  \bibinfo{author}{Postnov, D.~E.}
\newblock \bibinfo{journal}{\bibinfo{title}{Transport in the brain
  extracellular space: diffusion, but which kind?}}
\newblock {\emph{\JournalTitle{International Journal of Molecular Sciences}}}
  \textbf{\bibinfo{volume}{23}}, \bibinfo{pages}{12401} (\bibinfo{year}{2022}).

\bibitem{Tartakovsky2019}
\bibinfo{author}{Tartakovsky, D.~M.} \& \bibinfo{author}{Dentz, M.}
\newblock \bibinfo{journal}{\bibinfo{title}{Diffusion in porous media:
  phenomena and mechanisms}}.
\newblock {\emph{\JournalTitle{Transport in Porous Media}}}
  \textbf{\bibinfo{volume}{130}}, \bibinfo{pages}{105--127}
  (\bibinfo{year}{2019}).

\bibitem{Chen2000}
\bibinfo{author}{Chen, K.~C.} \& \bibinfo{author}{Nicholson, C.}
\newblock \bibinfo{journal}{\bibinfo{title}{Changes in brain cell shape create
  residual extracellular space volume and explain tortuosity behavior during
  osmotic challenge}}.
\newblock {\emph{\JournalTitle{Proceedings of the National Academy of
  Sciences}}} \textbf{\bibinfo{volume}{97}}, \bibinfo{pages}{8306--8311}
  (\bibinfo{year}{2000}).

\bibitem{Weber2022}
\bibinfo{author}{Weber, R.~M.}, \bibinfo{author}{Korneev, S.} \&
  \bibinfo{author}{Battiato, I.}
\newblock \bibinfo{journal}{\bibinfo{title}{Homogenization-informed
  convolutional neural networks for estimation of li-ion battery effective
  properties}}.
\newblock {\emph{\JournalTitle{Transport in Porous Media}}}
  \bibinfo{pages}{1--22} (\bibinfo{year}{2022}).

\bibitem{Wernert2022}
\bibinfo{author}{Wernert, V.} \emph{et~al.}
\newblock \bibinfo{journal}{\bibinfo{title}{Tortuosity of hierarchical porous
  materials: Diffusion experiments and random walk simulations}}.
\newblock {\emph{\JournalTitle{Chemical Engineering Science}}}
  \textbf{\bibinfo{volume}{264}}, \bibinfo{pages}{118136}
  (\bibinfo{year}{2022}).

\bibitem{Li2016}
\bibinfo{author}{Li, H.}, \bibinfo{author}{Li, H.}, \bibinfo{author}{Gao, B.},
  \bibinfo{author}{Wang, W.} \& \bibinfo{author}{Liu, C.}
\newblock \bibinfo{journal}{\bibinfo{title}{Study on pore characteristics and
  microstructure of sandstones with different grain sizes}}.
\newblock {\emph{\JournalTitle{Journal of Applied Geophysics}}}
  \textbf{\bibinfo{volume}{136}}, \bibinfo{pages}{364--371},
  \doiprefix\url{https://doi.org/10.1016/j.jappgeo.2016.11.015}
  (\bibinfo{year}{2017}).

\bibitem{Kinney2013}
\bibinfo{author}{Kinney, J.~P.} \emph{et~al.}
\newblock \bibinfo{journal}{\bibinfo{title}{Extracellular sheets and tunnels
  modulate glutamate diffusion in hippocampal neuropil}}.
\newblock {\emph{\JournalTitle{Journal of Comparative Neurology}}}
  \textbf{\bibinfo{volume}{521}}, \bibinfo{pages}{448--464}
  (\bibinfo{year}{2013}).

\bibitem{Godin2017}
\bibinfo{author}{Godin, A.~G.} \emph{et~al.}
\newblock \bibinfo{journal}{\bibinfo{title}{Single-nanotube tracking reveals
  the nanoscale organization of the extracellular space in the live brain}}.
\newblock {\emph{\JournalTitle{Nature Nanotechnology}}}
  \textbf{\bibinfo{volume}{12}}, \bibinfo{pages}{238--243},
  \doiprefix\url{10.1038/nnano.2016.248} (\bibinfo{year}{2017}).

\bibitem{Tartakovsky08}
\bibinfo{author}{Tartakovsky, A.~M.}, \bibinfo{author}{Tartakovsky, D.~M.} \&
  \bibinfo{author}{Meakin, P.}
\newblock \bibinfo{journal}{\bibinfo{title}{Stochastic langevin model for flow
  and transport in porous media}}.
\newblock {\emph{\JournalTitle{Phys. Rev. Lett.}}}
  \textbf{\bibinfo{volume}{101}}, \bibinfo{pages}{044502},
  \doiprefix\url{10.1103/PhysRevLett.101.044502} (\bibinfo{year}{2008}).

\bibitem{Kalz22}
\bibinfo{author}{Kalz, E.} \emph{et~al.}
\newblock \bibinfo{journal}{\bibinfo{title}{Collisions enhance self-diffusion
  in odd-diffusive systems}}.
\newblock {\emph{\JournalTitle{Phys. Rev. Lett.}}}
  \textbf{\bibinfo{volume}{129}}, \bibinfo{pages}{090601},
  \doiprefix\url{10.1103/PhysRevLett.129.090601} (\bibinfo{year}{2022}).

\bibitem{Alexandre22}
\bibinfo{author}{Alexandre, A.}, \bibinfo{author}{Mangeat, M.},
  \bibinfo{author}{Gu\'erin, T.} \& \bibinfo{author}{Dean, D.~S.}
\newblock \bibinfo{journal}{\bibinfo{title}{How stickiness can speed up
  diffusion in confined systems}}.
\newblock {\emph{\JournalTitle{Phys. Rev. Lett.}}}
  \textbf{\bibinfo{volume}{128}}, \bibinfo{pages}{210601},
  \doiprefix\url{10.1103/PhysRevLett.128.210601} (\bibinfo{year}{2022}).

\bibitem{Kasthuri2015}
\bibinfo{author}{Kasthuri, N.} \emph{et~al.}
\newblock \bibinfo{journal}{\bibinfo{title}{Saturated {{Reconstruction}} of a
  {{Volume}} of {{Neocortex}}}}.
\newblock {\emph{\JournalTitle{Cell}}} \textbf{\bibinfo{volume}{162}},
  \bibinfo{pages}{648--661}, \doiprefix\url{10.1016/j.cell.2015.06.054}
  (\bibinfo{year}{2015}).

\bibitem{Kamrava2020}
\bibinfo{author}{Kamrava, S.}, \bibinfo{author}{Tahmasebi, P.} \&
  \bibinfo{author}{Sahimi, M.}
\newblock \bibinfo{journal}{\bibinfo{title}{Linking morphology of porous media
  to their macroscopic permeability by deep learning}}.
\newblock {\emph{\JournalTitle{Transport in Porous Media}}}
  \textbf{\bibinfo{volume}{131}}, \doiprefix\url{10.1007/s11242-019-01352-5}
  (\bibinfo{year}{2020}).

\bibitem{Santos2021}
\bibinfo{author}{Santos, J.~E.} \emph{et~al.}
\newblock \bibinfo{journal}{\bibinfo{title}{Computationally efficient
  multiscale neural networks applied to fluid flow in complex 3d porous
  media}}.
\newblock {\emph{\JournalTitle{Transport in porous media}}}
  \textbf{\bibinfo{volume}{140}}, \bibinfo{pages}{241--272}
  (\bibinfo{year}{2021}).

\bibitem{Graczyk20}
\bibinfo{author}{Graczyk, K.~M.} \& \bibinfo{author}{Matyka, M.}
\newblock \bibinfo{journal}{\bibinfo{title}{Predicting porosity, permeability,
  and tortuosity of porous media from images by deep learning}}.
\newblock {\emph{\JournalTitle{Scientific reports}}}
  \textbf{\bibinfo{volume}{10}}, \bibinfo{pages}{1--11} (\bibinfo{year}{2020}).

\bibitem{Cawte2022}
\bibinfo{author}{Cawte, T.} \& \bibinfo{author}{Bazylak, A.}
\newblock \bibinfo{journal}{\bibinfo{title}{Accurately predicting transport
  properties of porous fibrous materials by machine learning methods}}.
\newblock {\emph{\JournalTitle{Electrochemical Science Advances}}}
  \bibinfo{pages}{e2100185} (\bibinfo{year}{2022}).

\bibitem{Sethi2022}
\bibinfo{author}{Ranjan~Sethi, S.}, \bibinfo{author}{Kumawat, V.} \&
  \bibinfo{author}{Ganguly, S.}
\newblock \bibinfo{journal}{\bibinfo{title}{Convolutional neural network based
  prediction of effective diffusivity from microscope images}}.
\newblock {\emph{\JournalTitle{Journal of Applied Physics}}}
  \textbf{\bibinfo{volume}{131}}, \bibinfo{pages}{214901}
  (\bibinfo{year}{2022}).

\bibitem{Roding2022}
\bibinfo{author}{R{\"o}ding, M.},
  \bibinfo{author}{W{\aa}hlstrand~Sk{\"a}rstr{\"o}m, V.} \&
  \bibinfo{author}{Lor{\'e}n, N.}
\newblock \bibinfo{journal}{\bibinfo{title}{Inverse design of anisotropic
  spinodoid materials with prescribed diffusivity}}.
\newblock {\emph{\JournalTitle{Scientific Reports}}}
  \textbf{\bibinfo{volume}{12}}, \bibinfo{pages}{17413},
  \doiprefix\url{10.1038/s41598-022-21451-6} (\bibinfo{year}{2022}).

\bibitem{Karmava2021c}
\bibinfo{author}{Kamrava, S.}, \bibinfo{author}{Tahmasebi, P.} \&
  \bibinfo{author}{Sahimi, M.}
\newblock \bibinfo{journal}{\bibinfo{title}{Physics- and image-based prediction
  of fluid flow and transport in complex porous membranes and materials by deep
  learning}}.
\newblock {\emph{\JournalTitle{Journal of Membrane Science}}}
  \textbf{\bibinfo{volume}{622}}, \bibinfo{pages}{119050},
  \doiprefix\url{https://doi.org/10.1016/j.memsci.2021.119050}
  (\bibinfo{year}{2021}).

\bibitem{Wu2018}
\bibinfo{author}{Wu, J.}, \bibinfo{author}{Yin, X.} \& \bibinfo{author}{Xiao,
  H.}
\newblock \bibinfo{journal}{\bibinfo{title}{Seeing permeability from images:
  fast prediction with convolutional neural networks}}.
\newblock {\emph{\JournalTitle{Science Bulletin}}}
  \textbf{\bibinfo{volume}{63}}, \bibinfo{pages}{1215--1222},
  \doiprefix\url{https://doi.org/10.1016/j.scib.2018.08.006}
  (\bibinfo{year}{2018}).

\bibitem{Kamrava2021}
\bibinfo{author}{Kamrava, S.}, \bibinfo{author}{Sahimi, M.} \&
  \bibinfo{author}{Tahmasebi, P.}
\newblock \bibinfo{journal}{\bibinfo{title}{Simulating fluid flow in complex
  porous materials by integrating the governing equations with deep-layered
  machines}}.
\newblock {\emph{\JournalTitle{npj Computational Materials}}}
  \textbf{\bibinfo{volume}{7}}, \bibinfo{pages}{127},
  \doiprefix\url{10.1038/s41524-021-00598-2} (\bibinfo{year}{2021}).

\bibitem{Tahmasebi2020}
\bibinfo{author}{Tahmasebi, P.}, \bibinfo{author}{Kamrava, S.},
  \bibinfo{author}{Bai, T.} \& \bibinfo{author}{Sahimi, M.}
\newblock \bibinfo{journal}{\bibinfo{title}{Machine learning in geo- and
  environmental sciences: From small to large scale}}.
\newblock {\emph{\JournalTitle{Advances in Water Resources}}}
  \textbf{\bibinfo{volume}{142}}, \bibinfo{pages}{103619},
  \doiprefix\url{https://doi.org/10.1016/j.advwatres.2020.103619}
  (\bibinfo{year}{2020}).

\bibitem{Graczyk2022}
\bibinfo{author}{Graczyk, K.~M.}, \bibinfo{author}{Paw{\l}owski, J.},
  \bibinfo{author}{Majchrowska, S.} \& \bibinfo{author}{Golan, T.}
\newblock \bibinfo{journal}{\bibinfo{title}{Self-normalized density map (sndm)
  for counting microbiological objects}}.
\newblock {\emph{\JournalTitle{Scientific Reports}}}
  \textbf{\bibinfo{volume}{12}}, \bibinfo{pages}{10583},
  \doiprefix\url{10.1038/s41598-022-14879-3} (\bibinfo{year}{2022}).

\bibitem{ronneberger2015unet}
\bibinfo{author}{Ronneberger, O.}, \bibinfo{author}{Fischer, P.} \&
  \bibinfo{author}{Brox, T.}
\newblock \bibinfo{title}{U-net: Convolutional networks for biomedical image
  segmentation} (\bibinfo{year}{2015}).
\newblock \eprint{1505.04597}.

\bibitem{gal2015dropout}
\bibinfo{author}{Gal, Y.} \& \bibinfo{author}{Ghahramani, Z.}
\newblock \bibinfo{title}{Dropout as a bayesian approximation: Representing
  model uncertainty in deep learning} (\bibinfo{year}{2015}).
\newblock \eprint{1506.02142}.

\bibitem{Wu2019}
\bibinfo{author}{Wu, H.}, \bibinfo{author}{Fang, W.-Z.}, \bibinfo{author}{Kang,
  Q.}, \bibinfo{author}{Tao, W.-Q.} \& \bibinfo{author}{Qiao, R.}
\newblock \bibinfo{journal}{\bibinfo{title}{Predicting effective diffusivity of
  porous media from images by deep learning}}.
\newblock {\emph{\JournalTitle{Scientific Reports}}}
  \textbf{\bibinfo{volume}{9}}, \bibinfo{pages}{20387},
  \doiprefix\url{10.1038/s41598-019-56309-x} (\bibinfo{year}{2019}).

\bibitem{NEURIPS2019_9015}
\bibinfo{author}{Paszke, A.} \emph{et~al.}
\newblock \bibinfo{title}{Pytorch: An imperative style, high-performance deep
  learning library}.
\newblock In \emph{\bibinfo{booktitle}{Advances in Neural Information
  Processing Systems 32}}, \bibinfo{pages}{8024--8035}
  (\bibinfo{publisher}{Curran Associates, Inc.}, \bibinfo{year}{2019}).

\bibitem{gal2015dropout-appendix}
\bibinfo{author}{Gal, Y.} \& \bibinfo{author}{Ghahramani, Z.}
\newblock \bibinfo{title}{Dropout as a bayesian approximation: Appendix}
  (\bibinfo{year}{2015}).
\newblock \eprint{1506.02157}.

\bibitem{Koza2014}
\bibinfo{author}{Koza, Z.}, \bibinfo{author}{Kondrat, G.} \&
  \bibinfo{author}{Suszczy{\'n}ski, K.}
\newblock \bibinfo{journal}{\bibinfo{title}{Percolation of overlapping squares
  or cubes on a lattice}}.
\newblock {\emph{\JournalTitle{Journal of Statistical Mechanics: Theory and
  Experiment}}} \textbf{\bibinfo{volume}{2014}}, \bibinfo{pages}{P11005}
  (\bibinfo{year}{2014}).

\bibitem{Boudreau1996}
\bibinfo{author}{Boudreau, B.~P.}
\newblock \bibinfo{journal}{\bibinfo{title}{The diffusive tortuosity of
  fine-grained unlithified sediments}}.
\newblock {\emph{\JournalTitle{Geochimica et Cosmochimica Acta}}}
  \textbf{\bibinfo{volume}{60}}, \bibinfo{pages}{3139--3142},
  \doiprefix\url{https://doi.org/10.1016/0016-7037(96)00158-5}
  (\bibinfo{year}{1996}).

\bibitem{Krueger2016}
\bibinfo{author}{Krüger, T.} \emph{et~al.}
\newblock \emph{\bibinfo{title}{The Lattice Boltzmann Method - Principles and
  Practice}} (\bibinfo{publisher}{Springer Cham}, \bibinfo{year}{2016}).

\bibitem{Succi2018}
\bibinfo{author}{Succi, S.}
\newblock \emph{\bibinfo{title}{The Lattice Boltzmann Equation: For Complex
  States of Flowing Matter}} (\bibinfo{publisher}{Oxford University Press},
  \bibinfo{year}{2018}).

\bibitem{Bhatnagar1954}
\bibinfo{author}{Bhatnagar, P.~L.}, \bibinfo{author}{Gross, E.~P.} \&
  \bibinfo{author}{Krook, M.}
\newblock \bibinfo{journal}{\bibinfo{title}{A {{Model}} for {{Collision
  Processes}} in {{Gases}}. {{I}}. {{Small Amplitude Processes}} in {{Charged}}
  and {{Neutral One-Component Systems}}}}.
\newblock {\emph{\JournalTitle{Phys. Rev.}}} \textbf{\bibinfo{volume}{94}},
  \bibinfo{pages}{511--525}, \doiprefix\url{10.1103/PhysRev.94.511}
  (\bibinfo{year}{1954}).

\end{thebibliography}
\section*{Author contributions statement}
K.M.G. D.S. and M.M. conceived the project. K.M.G. designed and performed the deep learning analysis and proposed usage of the CNN with the Self-Normalized module and U-Net architectures. D.S. designed and performed diffusion simulation within LBM and prepared the data. All authors participated in writing the text and discussing the results. MM found the exponents in Archie's law.

\section*{Additional information}
%

\textbf{Competing Interests:} The authors declare that they have no competing interests.


\end{document}